\documentclass[pra, preprint,showpacs,amsmath,amsfonts,amssymb,aps]
              {revtex4}
\usepackage{latexsym,makeidx,ifthen,calc,longtable,graphics}
\usepackage{bm}

\newcommand{\rqmversion}{Version 2}
\newcommand{\rqmtitle}{Spacetime Path Formalism for 
                       Massive Particles of Any Spin}
\newcommand{\rqmauthor}{Ed Seidewitz}
\newcommand{\rqmaddress}{14000 Gulliver's Trail, Bowie MD 20720 USA}

\begin{document}
    
    \title{\rqmtitle}
    \preprint{\rqmversion}
    
    \author{\rqmauthor}
    \email{seidewitz@mailaps.org}
    \affiliation{\rqmaddress}
    
    \date{\today}
    
    
    \begin{abstract}
        Earlier work presented a spacetime path formalism for
        relativistic quantum mechanics arising naturally from the
        fundamental principles of the Born probability rule,
        superposition, and spacetime translation invariance. The
        resulting formalism can be seen as a foundation for a number
        of previous parameterized approaches to relativistic quantum
        mechanics in the literature. Because time is treated similarly
        to the three space coordinates, rather than as an evolution
        parameter, such approaches have proved particularly useful in
        the study of quantum gravity and cosmology. The present paper
        extends the foundational spacetime path formalism to include
        massive, nonscalar particles of any (integer or half-integer)
        spin. This is done by generalizing the principle of
        translational invariance used in the scalar case to the
        principle of full Poincar\'e invariance, leading to a
        formulation for the nonscalar propagator in terms of a path
        integral over the Poincar\'e group. Once the difficulty of the
        non-compactness of the component Lorentz group is dealt with,
        the subsequent development is remarkably parallel to the
        scalar case. This allows the formalism to retain a clear
        probabilistic interpretation throughout, with a natural
        reduction to non-relativistic quantum mechanics closely
        related to the well known generalized Foldy-Wouthuysen
        transformation.
    \end{abstract}
    
    \maketitle
    
    \section{Introduction} \label{sect:intro}

Reference \onlinecite{seidewitz06a} presented a foundational formalism
for relativistic quantum mechanics based on path integrals over
parametrized paths in spacetime. As discussed there, such an approach
is particularly suited for further study of quantum gravity and
cosmology, and it can be given a natural interpretation in terms of
decoherent histories \cite{seidewitz06b}. However, the formalism as
given in \refcite{seidewitz06a} is limited to scalar particles. The
present paper extends this spacetime path formalism to non-scalar
particles, although the present work is still limited to massive
particles.

There have been several approaches proposed in the literature for
extending the path integral formulation of the relativistic scalar
propagator \cite{feynman50,feynman51,teitelboim82,hartle92} to the
case of non-scalar particles, particularly spin-1/2 (see, for example,
\refcites{bordi80, henneaux82, barut84, mannheim85, forte05}). These
approaches generally proceed by including in the path integral
additional variables to represent higher spin degrees of freedom.
However, there is still a lack of a comprehensive path integral
formalism that treats all spin values in a consistent way, in the
spirit of the classic work of Weinburg \cite{weinberg64a, weinberg64b,
weinberg69} for traditional quantum field theory. Further, most
earlier references assume that the path integral approach is basically
a reformulation of an \emph{a priori} traditional Hamiltonian
formulation of quantum mechanics, rather than being foundational in
its own right.

The approach to be considered here extends the approach from
\refcite{seidewitz06a} to non-scalar particles by expanding the
configuration space of a particle to be the Poincar\'{e} group (also
known as the inhomogeneous Lorentz group). That is, rather than just
considering the position of a particle, the configuration of a
particle will be taken to be both a position \emph{and} a Lorentz
transformation. Choosing various representations of the group of
Lorentz transformations then allows all spins to be handled in a
consistent way.

The idea of using a Lorentz group variable to represent spin degrees
of freedom is not new. For example, Hanson and Regge \cite{hanson74}
describe the physical configuration of a relativistic spherical top as
a Poincar\'e element whose degrees of freedom are then restricted.
Similarly, Hannibal \cite{hannibal97} proposes a full canonical
formalism for classical spinning particles using the Lorentz group for
the spin configuration space, which is then quantized to describe both
spin and isospin. Rivas \cite{rivas89, rivas94, rivas01} has made a
comprehensive study in which an elementary particle is defined as ``a
mechanical system whose kinematical space is a homogeneous space of
the Poincar\'e group''.

Rivas actually proposes quantization using path integrals, but he does
not provide an explicit derivation of the non-scalar propagator by
evaluating such an integral. A primary goal of this paper to provide
such a derivation.

Following a similar approach to \refcite{seidewitz06a}, the form of
the path integral for non-scalar particles will be deduced from the
fundamental principles of the Born probability rule, superposition,
and Poincar\'e invariance. After a brief overview in
\sect{sect:background} of some background for this approach,
\sect{sect:non-scalar:propagator} generalizes the postulates from
\refcite{seidewitz06a} to the non-scalar case, leading to a path
integral over an appropriate Lagrangian function on the Poincar\'e
group variables.

The major difficulty with evaluating this path integral is the
non-compactness of the Lorentz group. Previous work on evaluating
Lorentz group path integrals (going back to \refcite{bohm87}) is based
on the irreducible unitary representations of the group. This is
awkward, since, for a non-compact group, these representations are
continuous \cite{vilenkin68} and the results do not generalize easily
to the covering group $SL(2,\cmplx)$ that includes half-integral
spins.

Instead, we will proceed by considering a Wick rotation to Euclidean
space, which replaces the non-compact Lorentz group $SO(3,1)$ by the
compact group $SO(4)$ of rotations in four dimensions, in which it is
straightforward to evaluate the path integral. It will then be argued
that, even though the $SO(4)$ propagator cannot be assumed the same as
the true Lorentz propagator, the propagators should be the same when
restricted to the common subgroup $SO(3)$ of rotations in three
dimensions. This leads directly to considerations of the spin
representations of $SO(3)$.

Accordingly, \sect{sect:non-scalar:euclidean} develops the Euclidean
$SO(4)$ propagator and \sect{sect:non-scalar:spin} then considers the
reduction to the three-dimensional rotation group and its spin
representations. However, rather than using the usual Wigner approach
of reduction along the momentum vector \cite{wigner39}, we will reduce
along an independent time-like four-vector \cite{piron78, horwitz82}.
This allows for a very parallel development to \refcite{seidewitz06a}
for antiparticles in \sect{sect:non-scalar:antiparticles} and for a
clear probability interpretation in
\sect{sect:non-scalar:probability}.

Interactions of non-scalar particles can be included in the formalism
by a straightforward generalization of the approach given in
\refcite{seidewitz06a}. \Sect{sect:non-scalar:interactions} gives an 
overview of this, though full details are not included where they are 
substantially the same as the scalar case.

Natural units with $\hbar = 1 = c$ are used throughout the following 
and the metric has a signature of $(- + + +)$.

\section{Background} 
\label{sect:background}

Path integrals were originally introduced by Feynman \cite{feynman48,
feynman65} to represent the non-relativistic propagation kernel
$\kersym(\threex_{1} - \threex_{0}; t_{1}-t_{0})$. This kernel gives the
transition amplitude for a particle to propagate from the position
$\threex_{0}$ at time $t_{0}$ to the position $\threex_{1}$ at time
$t_{1}$. That is, if $\psi(\threex_{0}; t_{0})$ is the probability 
amplitude for the particle to be at position $\threex_{0}$ at time 
$t_{0}$, then the amplitude for it to propagate to another position 
at a later time is
\begin{equation*}
    \psi(\threex; t) = \intthree \xz\, 
        \kersym(\threex - \threex_{0}; t-t_{0}) 
        \psi(\threex_{0}; t_{0}) \,.
\end{equation*}

A specific \emph{path} of a particle in space is given by a
position function $\threevec{q}(t)$ parametrized by time (or, in
coordinate form, the three functions $q^{i}(t)$ for $i = 1,2,3$). 
Now consider all possible paths starting at $\threevec{q}(t_{0}) = 
\threex_{0}$ and ending at $\threevec{q}(t_{1}) = \threex_{1}$. The 
path integral form for the propagation kernel is then given by 
integrating over all these paths as follows:
\begin{equation} \label{eqn:A0a}
    \kersym(\threex_{1} - \threex_{0}; t_{1}-t_{0})
        = \zeta \intDthree q\, 
              \delta^{3}(\threevec{q}(t_{1}) - \threex_{1})
              \delta^{3}(\threevec{q}(t_{0}) - \threex_{0})
              \me^{\mi S[\threevec{q}]} \,,
\end{equation}
where the phase function $S[\threevec{q}]$ is given by the classical 
action
\begin{equation*}
    S[\threevec{q}] \equiv 
        \int_{t_{0}}^{t_{1}} \dt\, L(\dot{\threevec{q}}(t)) \,,
\end{equation*}
with $L(\dot{\threevec{q}})$ being the non-relativistic Lagrangian in
terms of the three-velocity $\dot{\threevec{q}} \equiv
\dif\threevec{q} / \dt$.

In \eqn{eqn:A0a}, the notation $\Dthree q$ indicates a path integral
over the three functions $q^{i}(t)$. The Dirac delta functions
constrain the paths integrated over to start and end at the
appropriate positions. Finally, $\zeta$ is a normalization factor,
including any limiting factors required to keep the path integral
finite (which are sometimes incorporated into the integration measure
$\Dthree q$ instead).

As later noted by Feynman himself \cite{feynman51}, it is possible to
generalize the path integral approach to the relativistic case. To do
this, it is necessary to consider paths in \emph{spacetime}, rather
than just space. Such a path is given by a four dimensional position
function $q(\lambda)$, parametrized by an invariant \emph{path
parameter} $\lambda$ (or, in coordinate form, the four functions
$\qmul$, for $\mu = 0,1,2,3$).

The propagation amplitude for a free scalar particle in spacetime is 
given by the Feynman propagator
\begin{equation} \label{eqn:A0b}
    \prop = -\mi(2\pi)^{-4}\intfour p\, 
                \frac{\me^{\mi p\cdot(x-\xz)}}
                     {p^{2}+m^{2}-\mi\epsilon} \,.
\end{equation}
It can be shown (in addition to \refcite{feynman51}, see also, e.g.,
\refcites{seidewitz06a, teitelboim82}) that this propagator can be
expressed in path integral form as
\begin{equation} \label{eqn:A0c}
    \prop = \int_{\lambdaz}^{\infty} \dif\lambda\,
        \zeta \intDfour q\,
        \delta^{4}(q(\lambda) - x) \delta^{4}(q(\lambdaz) - \xz)
        \me^{\mi S[q]} \,,
\end{equation}
where
\begin{equation*}
    S[q] \equiv 
        \int_{\lambdaz}^{\lambda} \dif\lambda'\, L(\qdot)(\lambda')) \,,
\end{equation*}
and $L(\qdot)$ is now the relativistic Lagrangian in terms of the the 
four-velocity $\qdot \equiv \dif q / \dl$.

Notice that the form of the relativistic expression differs from the 
non-relativistic one by having an additional integration over 
$\lambda$. This is necessary, since the propagator must, in the end, 
depend only on the change in position, independent of $\lambda$. 
However, as noted in \refcite{seidewitz06a}, \eqn{eqn:A0c} can be 
written as
\begin{equation} \label{eqn:A0d}
    \prop = \int_{\lambdaz}^{\infty} \dif\lambda\, \kerneld \,,
\end{equation}
where the \emph{relativistic kernel}
\begin{equation} \label{eqn:A0e}
    \kerneld = \zeta \intDfour q\,
        \delta^{4}(q(\lambda) - x) \delta^{4}(q(\lambdaz) - \xz)
        \me^{\mi S[q]}
\end{equation}
now has a form entirely parallel with the non-relativistic case. The
relativistic kernel can be considered to represent propagation over
paths of the specific length $\lambda - \lambdaz$, while \eqn{eqn:A0d}
then integrates over all possible path lengths.

Given the parallel with the non-relativistic case, define the
\emph{parametrized} probability amplitudes $\psixl$ such that
\begin{equation*}
    \psixl = \intfour \xz\, \kerneld \psixlz \,.
\end{equation*}
Parametrized amplitudes were introduced by Stueckelberg
\cite{stueckelberg41, stueckelberg42}, and parametrized approaches to
relativistic quantum mechanics have been developed by a number of
subsequent authors \cite{nambu50, schwinger51, cooke68, horwitz73,
collins78, piron78, fanchi78, fanchi83, fanchi93}. The approach is
developed further in the context of spacetime paths of scalar
particles in \refcite{seidewitz06a}.

In the traditional presentation, however, it is not at all clear
\emph{why} the path integrals of \eqns{eqn:A0a} and \eqref{eqn:A0b}
should reproduce the expected results for non-relativistic and
relativistic propagation. The phase functional $S$ is simply chosen to
have the form of the classical action, such that this works. In
contrast, \refcite{seidewitz06a} makes a more fundamental argument
that the exponential form of \eqn{eqn:A0e} is a consequence of
translation invariance in Minkowski spacetime. This allows for
development of the spacetime path formalism as a foundational
approach, rather than just a re-expression of already known results.

The full invariant group of Minkowski spacetime is not the
translation group, though, but the Poincar\'e group consisting of both
translations \emph{and} Lorentz transformations. This leads one to
consider the implications of applying the argument of
\refcite{seidewitz06a} to the full Poincar\'e group.

Now, while a translation applies to the position of a particle, a
Lorentz transformation applies to its \emph{frame of reference}. Just
as we can consider the position $x$ of a particle to be a translation
by $x$ from some fixed origin $O$, we can consider the frame of
reference of a particle to be given by a Lorentz transformation
$\Lambda$ from a fixed initial frame $I$. The full configuration of a
particle is then given by $(x,\Lambda)$, for a position $x$ and a
Lorentz transformation $\Lambda$---that is, the configuration space of
the particle is also just the Poincar\'e group. The application of an
arbitrary Poincar\'e transformation $(\Delta x,\Lambda')$ to a
particle configuration $(x,\Lambda)$ results in the transformed
configuration $(\Lambda' x + \Delta x, \Lambda' \Lambda)$.

A particle path will now be a path through the Poincar\'e group, not
just through spacetime. Such a path is given by both a position
function $q(\lambda)$ \emph{and} a Lorentz transformation function
$M(\lambda)$ (in coordinate form, a Lorentz transformation is
represented by a matrix, so there are \emph{sixteen} functions
$\hilo{M}{\mu}{\nu}(\lambda)$, for $\mu,\nu = 0,1,2,3,$). The
remainder of this paper will re-develop the spacetime path formalism
introduced in \refcite{seidewitz06a} in terms of this expanded
conception of particle paths. As we will see, this naturally leads to
a model for non-scalar particles.

\section{The Non-scalar Propagator} 
\label{sect:non-scalar:propagator}

This section develops the path-integral form of the non-scalar
propagator from the conception of Poincar\'e group particle paths
introduced in the previous section. The argument parallels that of 
\refcite{seidewitz06a} for the scalar case, motivating a set of 
postulates that lead to the appropriate path integral form.

To begin, let $\kersym(x-\xz, \Lambda\Lambdaz^{-1}; \lambda-\lambdaz)$
be the transition amplitude for a particle to go from the
configuration $(\xz, \Lambdaz)$ at $\lambdaz$ to the configuration
$(x, \Lambda)$ at $\lambda$. By Poincar\'e invariance, this amplitude
only depends on the relative quantities $x-\xz$ and
$\Lambda\Lambdaz^{-1}$. By parameter shift invariance, it only depends
on $\lambda-\lambdaz$. Similarly to the scalar case (\eqn{eqn:A0d}),
the full propagator is given by integrating over the kernel path
length parameter:
\begin{equation} \label{eqn:A1a} 
    \propsym(x-\xz,\Lambda\Lambdaz^{-1})
        = \int_{0}^{\infty} \dl\, 
                       \kersym(x-\xz,\Lambda\Lambdaz^{-1};\lambda) \,.
\end{equation}

The fundamental postulate of the spacetime path approach is that a
particle's transition amplitude between two points is a superposition
of the transition amplitudes for all possible paths between those
points. Let the functional $\propsym[q, M]$ give the transition
amplitude for a path $q(\lambda), M(\lambda)$. Then the transition
amplitude $\kersym(x-\xz, \Lambda\Lambdaz^{-1}; \lambda-\lambdaz)$
must be given by a path integral over $\propsym[q, M]$ for all paths
starting at $(\xz, \Lambdaz)$ and ending at $(x, \Lambda)$ with the
parameter interval $[\lambdaz, \lambda]$.

\begin{postulate}
    For a free, non-scalar particle, the transition amplitude
    $\kersym(x-\xz, \Lambda\Lambdaz^{-1}; \lambda-\lambdaz)$ is given
    by the superposition of path transition amplitudes $\propsym[q,
    M]$, for all possible Poincar\'e path functions $q(\lambda),
    M(\lambda)$ beginning at $(\xz, \Lambdaz)$ and ending at $(x,
    \Lambda)$, parametrized over the interval $[\lambdaz, \lambda]$.
    That is,
    \begin{multline} \label{eqn:A1} 
        \kersym(x-\xz, \Lambda\Lambdaz^{-1}; \lambda-\lambdaz)
            = \zeta \intDfour q\, \intDsix M\, \\
                \delta^{4}(q(\lambda) - x) 
                \delta^{6}(M(\lambda)\Lambda^{-1} - I)
                \delta^{4}(q(\lambdaz) - \xz)
                \delta^{6}(M(\lambdaz)\Lambdaz^{-1} - I)
                \propsym[q, M] \,,
    \end{multline}
    where $\zeta$ is a normalization factor as required to keep the
    path integral finite.
\end{postulate}

As previously noted, the notation $\Dfour q$ in \eqn{eqn:A1} indicates
a path integral over the four path functions $\qmul$. Similarly,
$\Dsix M$ indicates a path integral over the Lorentz group functions
$\hilo{M}{\mu}{\nu}(\lambda)$. While a Lorentz transformation matrix
$[\hilo{\Lambda}{\mu}{\nu}]$ has sixteen elements, any such matrix is
constrained by the condition
\begin{equation} \label{eqn:A2}
    \hilo{\Lambda}{\alpha}{\mu}\eta_{\alpha\beta}
        \hilo{\Lambda}{\beta}{\nu} = \eta_{\mu\nu} \,,
\end{equation}
where $[\eta_{\mu\nu}]=\mathrm{diag}(-1,1,1,1)$ is the flat Minkowski
space metric tensor. This equation is symmetric, so it introduces ten
constraints, leaving only six actual degrees of freedom for a Lorentz
transformation. The Lorentz group is thus six dimensional, as
indicated by the notation $\Dsix$ in the path integral.

To further deduce the form of $\propsym[q, M]$, consider a family of
particle paths $q_{\xz,\Lambdaz}, M_{\xz,\Lambdaz}$, indexed by the
starting configuration $(\xz, \Lambdaz)$, such that
\begin{equation*}
    q_{\xz,\Lambdaz}(\lambda) 
        = \xz + \Lambdaz \tilde{q}(\lambda)
    \quad \text{and} \quad
    M_{\xz,\Lambdaz}(\lambda) = \Lambdaz \tilde{M}(\lambda) \,,
\end{equation*}
where $\tilde{q}(\lambdaz) = 0$ and $\tilde{M}(\lambdaz) = I$. These
paths are constructed by, in effect, applying the Poincar\'e
transformation given by $(\xz, \Lambdaz)$ to the specific functions
$\tilde{q}$ and $\tilde{M}$ defining the family. (Note how the 
ability to do this depends on the particle configuration space being 
the same as the Poincar\'e transformation group.)

Consider, though, that the particle propagation embodied in
$\kersym[q,M]$ must be Poincar\'e invariant. That is, $\kersym[q',M']
= \kersym[q,M]$ for any $q',M'$ related to $q,M$ by a fixed Poincar\'e
transformation. Thus, all members of the family $q_{\xz,\Lambdaz},
M_{\xz,\Lambdaz}$, which are all related to $\tilde{q}. \tilde{M}$ by
Poincar\'e transformations, must have the same amplitude
$\kersym[q_{\xz,\Lambdaz}, M_{\xz,\Lambdaz}] = \kersym[\tilde{q},
\tilde{M}]$, depending only on the functions $\tilde{q}$ and
$\tilde{M}$.

Suppose that a probability amplitude $\psi(\xz, \Lambdaz)$ is given
for a particle to be at in an initial configuration $(\xz,\Lambdaz)$
and that the transition amplitude is known to be $\kersym[\tilde{q},
\tilde{M}]$ for specific relative configuration functions $\tilde{q},
\tilde{M}$. Then, the probability amplitude for the particle to
traverse a specific path $(q_{\xz,\Lambdaz}(\lambda),
M_{\xz,\Lambdaz}(\lambda))$ from the family defined by the functions
$\tilde{q}, \tilde{M}$ should be just $\kersym[q_{\xz,\Lambdaz},
M_{\xz,\Lambdaz}] \psi(\xz, \Lambdaz) = \kersym[\tilde{q}, \tilde{M}]
\psi(\xz, \Lambdaz)$.

However, the very meaning of being on a specific path is that the
particle must propagate from the given starting configuration to the
specific ending configuration of the path. Further, since the paths in
the family are parallel in configuration space, the ending
configuration is uniquely determined by the starting configuration.
Therefore, the probability for reaching the ending configuration must
be the same as the probability for having started out at the given
initial configuration $(\xz,\Lambdaz)$. That is,
\begin{equation*}
    \sqr{\kersym[\tilde{q}, \tilde{M}]\psi(\xz,\Lambdaz)}
        = \sqr{\psi(\xz,\Lambdaz)} \,.
\end{equation*}
But, since $\kersym[\tilde{q}, \tilde{M}]$ is independent of $\xz$ and
$\Lambdaz$, we must have $\sqr{\kersym[q, M]} = 1$ in general.

This argument therefore suggests the following postulate.

\begin{postulate}
    For any path $(q(\lambda),M(\lambda))$, the transition amplitude
    $\propsym[q,M]$ preserves the probability density for the particle
    along the path. That is, it satisfies
    \begin{equation} \label{eqn:A3}
        \sqr{\propsym[q,M]} = 1 \,.
    \end{equation}    
\end{postulate}

The requirements of \eqn{eqn:A3} and Poincar\'e invariance mean that 
$\propsym[q,M]$ must have the exponential form
\begin{equation} \label{eqn:A4}
    \propsym[q,M] = \me^{\mi S[\tilde{q}, \tilde{M}]} \,,
\end{equation}
for some phase functional $S$ of the \emph{relative} path functions
\begin{equation*}
    \tilde{q}(\lambda) \equiv M(\lambdaz)^{-1}(q(\lambda)-q(\lambdaz))
    \quad \text{and} \quad
    \tilde{M}(\lambda) \equiv M(\lambdaz)^{-1}M(\lambda) \,.
\end{equation*}

As discussed in \refcite{seidewitz06a}, we are actually justified in 
replacing these relative functions with path derivatives under the 
path integral, even though the path functions $q(\lambda)$ and 
$M(\lambda)$ may not themselves be differentiable in general. This is 
because a path integral is defined as the limit of discretized 
approximations in which path derivatives are approximated as mean 
values, and the limit is then taken over the path integral as a 
whole, not each derivative individually. Thus, even though the 
individual path derivative limits may not be defined, the path 
integral has a well-defined value so long as the overall path 
integral limit is defined.

However, the quantities $\tilde{q}$ and $\tilde{M}$ are expressed in a
frame that varies with the $M(\lambdaz)$ of the specific path under
consideration. We wish instead to construct differentials in the fixed
``laboratory'' frame of the $q(\lambda)$. Transforming $\tilde{q}$ and
$\tilde{M}$ to this frame gives
\begin{equation*}
     M(\lambdaz)\tilde{q}(\lambda) = q(\lambda)-q(\lambdaz)
    \quad \text{and} \quad
    M(\lambdaz)\tilde{M}(\lambda)M(\lambdaz)^{-1} 
        = M(\lambda)M(\lambdaz)^{-1} \,.
\end{equation*}

Clearly, the corresponding derivative for $q$ is simply
$\qdot(\lambda) \equiv \dif q / \dl$, which is the tangent vector to
the path $q(\lambda)$. The derivative for $M$ needs to be treated a
little more carefully. Since the Lorentz group is a \emph{Lie group}
(that is, a continuous, differentiable group), the tangent to a path
$M(\lambda)$ in the Lorentz group space is given by an element of the
corresponding \emph{Lie algebra} \cite{warner83, frankel97}. For the
Lorentz group, the proper such tangent is given by the matrix
$\Omega(\lambda) = \dot{M}(\lambda)M(\lambda)^{-1}$, where
$\dot{M}(\lambda) \equiv \dif M / \dl$.

Together, the quantities $(\qdot, \Omega)$ form a tangent along the
path in the full Poincar\'e group space. We can then take the
arguments of the phase functional in \eqn{eqn:A4} to be $(\qdot,
\Omega)$. Substituting this into \eqn{eqn:A1} gives
\begin{multline} \label{eqn:A5}
    \kersym(x-\xz, \Lambda\Lambdaz^{-1}; \lambda-\lambdaz)
        = \zeta \intDfour q\, \intDsix M\, \\
            \delta^{4}(q(\lambda) - x) 
            \delta^{6}(M(\lambda)\Lambda^{-1} - I)
            \delta^{4}(q(\lambdaz) - \xz)
            \delta^{6}(M(\lambdaz)\Lambdaz^{-1} - I)
            \me^{\mi S[\qdot, \Omega]} \,,
\end{multline}
which reflects the typical form of a Feynman sum over paths.

Now, by dividing a path $(q(\lambda), M(\lambda))$ into two paths at 
some arbitrary parameter value $\lambda$ and propagating over each 
segment, one can see that
\begin{equation}
    S[\qdot, \Omega;\lambda_{1},\lambdaz] 
        = S[\qdot, \Omega;\lambda_{1},\lambda] +
    Ê     S[\qdot, \Omega;\lambda,\lambdaz] \,,
\end{equation}
where $S[\qdot, \Omega;\lambda',\lambda]$ denotes the value of
$S[\qdot,\Omega]$ for the path parameter range restricted to
$[\lambda,\lambda']$. Using this property to build the total value of
$S[\qdot,\Omega]$ from infinitesimal increments leads to the following result
(whose full proof is a straightforward generalization of the proof 
given in \refcite{seidewitz06a} for the scalar case).

\begin{theorem}[Form of the Phase Functional]
    The phase functional $S$ must have the form
    \begin{equation*} 
        S[\qdot,\Omega] = \int^{\lambda_{1}}_{\lambdaz} \dl' \, 
                          L[\qdot,\Omega;\lambda'] \,,
    \end{equation*}
    where the parametrization domain is $[\lambdaz,\lambda_{1}]$ and
    $L[\qdot, \Omega;\lambda]$ depends only on $\qdot$, $\Omega$ and
    their higher derivatives evaluated at $\lambda$.
\end{theorem}

Clearly, the functional $L[\qdot,\Omega; \lambda]$ plays the traditional role 
of the Lagrangian. The simplest non-trivial form for this functional 
would be for it to depend only on $\qdot$ and $\Omega$ and no higher 
derivatives. Further, suppose that it separates into uncoupled parts 
dependent on $\qdot$ and $\Omega$: 
\begin{equation*}
    L[\qdot,\Omega; \lambda] = 
        L_{q}[\qdot;\lambda] + L_{M}[\Omega; \lambda] \,.
\end{equation*}
The path integral of \eqn{eqn:A5} then factors into independent parts 
in $q$ and $M$, such that
\begin{equation} \label{eqn:A5a}
    \kersym(x-\xz, \Lambda\Lambdaz^{-1}; \lambda-\lambdaz) =
        \kerneld \kersym(\Lambda\Lambdaz^{-1}; \lambda-\lambdaz) \,.
\end{equation}
If we take $L_{q}$ to have the classical Lagrangian form
\begin{equation*}
    L_{q}[\qdot;\lambda] = L_{q}(\qdot(\lambda))
        = \frac{1}{4}\qdot(\lambda)^{2} - m^{2} \,,
\end{equation*}
for a particle of mass $m$, then the path integral in $q$ can be 
evaluated to give \cite{seidewitz06a,teitelboim82}
\begin{equation} \label{eqn:A5b}
    \kerneld = (2\pi)^{-4}\intfour p\, \me^{ip\cdot(x-\xz)}
                  \me^{-\mi(\lambda-\lambdaz)(p^{2}+m^{2})} \,.
\end{equation}
Similarly, take $L_{M}$ to be a Lorentz-invariant scalar function of
$\Omega(\lambda)$. $\Omega$ is an antisymmetric matrix (this can be
shown by differentiating the constraint \eqn{eqn:A2}), so the scalar
$\tr(\Omega) = \hilo{\Omega}{\mu}{\mu} = 0$. The next simplest choice
is
\begin{equation*}
    L_{M}[\Omega;\lambda] = L_{M}(\Omega(\lambda))
        = \frac{1}{2}\tr(\Omega(\lambda)\Omega(\lambda)\T)
        = \frac{1}{2}\Omega^{\mu\nu}(\lambda)
                     \Omega_{\mu\nu}(\lambda) \,.
\end{equation*}

\begin{postulate}
    For a free non-scalar particle of mass $m$, the Lagrangian 
    function is given by
    \begin{equation*}
        L(\qdot,\Omega) = L_{q}(\qdot) + L_{M}(\Omega) \,,
    \end{equation*}
    where
    \begin{equation*}
        L_{q}(\qdot) = \frac{1}{4}\qdot^{2} - m^{2}
    \end{equation*}
    and
    \begin{equation*}
        L_{M}(\Omega) = \frac{1}{2}\tr(\Omega\Omega\T) \,.
    \end{equation*}
\end{postulate}

Evaluating the path integral in $M$ is complicated by the fact that
the Lorentz group is not \emph{compact}, and integration over the
group is not, in general, bounded. The Lorentz group is denoted
$SO(3,1)$ for the three plus and one minus sign of the Minkowski
metric $\eta$ in the defining pseudo-orthogonality condition
\eqn{eqn:A2}. It is the minus sign on the time component of $\eta$
that leads to the characteristic Lorentz boosts of special relativity.
But since such boosts are parametrized by the boost velocity,
integration of this sector of the Lorentz group is unbounded. This is
in contrast to the three dimensional rotation subgroup $SO(3)$ for the
Lorentz, which is parameterized by rotation angles that are bounded.

To avoid this problem, we will \emph{Wick rotate} \cite{wick50} the
time axis in complex space. This replaces the physical $t$ coordinate
with $\mi t$, turning the minus sign in the metric to a plus sign,
resulting in the normal Euclidean metric $\mathrm{diag}(1,1,1,1)$. The
symmetry group of Lorentz transformations in Minkowski space then
corresponds to the symmetry group $SO(4)$ of rotations in
four-dimensional Euclidean space. The group $SO(4)$ \emph{is} compact,
and the path integration over $SO(4)$ can be done \cite{bohm87}.

Rather than dividing into boost and rotational parts, like the Lorentz
group, $SO(4)$ instead divides into two $SO(3)$ subgroups of rotations
in three dimensions. Actually, rather than $SO(3)$ itself, it is more
useful to consider its universal covering group $SU(2)$, the group of
two-dimensional unitary matrices, because $SU(2)$ allows for
representations with half-integral spin \cite{weyl50, weinberg95,
frankel97}. (The covering group $SU(2) \times SU(2)$ for $SO(4)$ in
Euclidean space corresponds to the covering group $SL(2,\cmplx)$ of
two-dimensional complex matrices for the Lorentz group $SO(3,1)$ in
Minkowski space.)

Typically, Wick rotations have been used to simplify the evaluation of
path integrals parametrized in time, like the non-relativistic
integral of \eqn{eqn:A0a}. In this case, replacing $t$ by $\mi t$
results in the exponent in the integrand of the path integral to
become real. Unlike this case, the exponent in the integrand of a
spacetime path integral remains imaginary, since the Wick rotation
does not affect the path parameter $\lambda$. Nevertheless, the path
integral can be evaluated, giving the following result (proved in the
Appendix).

\begin{theorem}[Evaluation of the $SO(4)$ Path Integral]
    Consider the path integral
    \begin{multline} \label{eqn:A6a}
        \kersym(\LambdaE\LambdaEz^{-1};\lambda-\lambdaz)
            = \euc{\eta} \intDsix \ME\,
                \delta^{6}(\ME(\lambda)\LambdaE^{-1}-I)
                \delta^{6}(\ME(\lambdaz)\LambdaEz^{-1}-I) \\
                \exp \left[
                    \mi\int^{\lambda}_{\lambdaz}\dl'\,
                        \frac{1}{2} \tr(\OmegaE(\lambda')
                                        \OmegaE(\lambda')\T)
                \right]
    \end{multline}
    over the six dimensional group $SO(4) \sim SU(2) \times SU(2)$,
    where $\OmegaE(\lambda')$ is the element of the Lie algebra $so(4)$
    tangent to the path $\ME(\lambda)$ at $\lambda'$. This path integral
    may be evaluated to get
    \begin{multline} \label{eqn:A6}
        \kersym(\LambdaE\LambdaEz^{-1};\lambda-\lambdaz) \\
            = \sum_{\ell_{A},\ell_{B}}
                \exp^{-\mi( \Delta m_{\ell_{A}}^{2} 
                          + \Delta m_{\ell_{B}}^{2})
                          (\lambda - \lambdaz)}
                (2\ell_{A}+1)(2\ell_{B}+1)
                \chi^{(\ell_{A}\ell_{B})}(\LambdaE\LambdaEz^{-1}) \,,
    \end{multline}
    where the summation over $\ell_{A}$ and $\ell_{B}$ is from $0$ to
    $\infty$ in steps of $1/2$, $\Delta m_{\ell}^{2} = \ell(\ell+1)$
    and $\chi^{(\ell_{A},\ell_{B})}$ is the group character for the
    $(\ell_{A},\ell_{B})$ $SU(2) \times SU(2)$ group representation.
\end{theorem}

The result of \eqn{eqn:A6} is in terms of the \emph{representations}
of the covering group $SU(2) \times SU(2)$. A (matrix) representation
$L$ of a group assigns to each group element $g$ a matrix $D^{(L)}(g)$
that respects the group operation, that is, such that
$D^{(L)}(g_{1}g_{2}) = D^{(L)}(g_{1}) D^{(L)}(g_{2})$. The
\emph{character function} $\chi^{(L)}$ for the representation $L$ of a
group is a function from the group to the reals such that
\begin{equation*}
    \chi^{(L)}(g) \equiv \tr(D^{(L)}(g)) \,.
\end{equation*}

The group $SU(2)$ has the well known \emph{spin representations},
labeled by spins $\ell = 0, 1/2, 1, 3/2, \ldots$ \cite{weyl50,
weinberg95} (for example, spin 0 is the trivial scalar representation,
spin 1/2 is the spinor representation and spin 1 is the vector
representation). A $(\ell_{A},\ell_{B})$ representation of $SU(2)
\times SU(2)$ then corresponds to a spin-$\ell_{A}$ representation for
the first $SU(2)$ component and a spin-$\ell_{B}$ representation for
the second $SU(2)$ component.

Of course, it is not immediately clear that this result for $SO(4)$
applies directly to $SO(3,1)$. In some cases, it can be shown that the
evolution propagator for a non-compact group is, in fact, the same as
the propagator for a related compact group. Unfortunately, the
relationship between $SO(4)$ and $SO(3,1)$ (in which an odd number,
three, of the six generators of $SO(4)$ are multiplied by $\mi$ to get
the boost generators for $SO(3,1)$) is such that the evolution
propagator of the non-compact group does not coincide with that of the
compact group \cite{krausz00}.

Nevertheless, $SO(4)$ and $SO(3,1)$ both have compact $SO(3)$
subgroups, which are isomorphic. Therefore, the restriction of the
$SO(4)$ propagator to its $SO(3)$ subgroup should correspond to the
restriction of the $SO(3,1)$ propagator to its $SO(3)$ subgroup. This
will prove sufficient for our purposes. In the next section, we will
continue to freely work with the Wick rotated Euclidean space and the
$SO(4)$ propagator as necessary. To show clearly when this is being
done, quantities effected by Wick rotation will be given a subscript
$E$, as in \eqn{eqn:A6}.
        
\section{The Euclidean Propagator} \label{sect:non-scalar:euclidean}

For a scalar particle, one can define the probability amplitude
$\psixl$ for the particle to be at position $x$ at the point $\lambda$
in its path \cite{seidewitz06a, stueckelberg41, stueckelberg42}. For a
non-scalar particle, this can be extended to a probability amplitude
$\psixLl$ for the particle to be in the Poincar\'{e} configuration
$(x,\Lambda)$, at the point $\lambda$ in its path. The transition
amplitude given in \eqn{eqn:A1} acts as a propagation kernel for
$\psixLl$:
\begin{equation*} 
    \psixLl = \intfour \xz\, \intsix \Lambdaz\, 
                  \kersym(x-\xz,\Lambda\Lambdaz^{-1};\lambda-\lambdaz) 
                  \psilz{\xz, \Lambdaz} \,.
\end{equation*}
The Euclidean version of this equation has an identical form, but in
terms of Euclidean configuration space quantities:
\begin{equation} \label{eqn:B1} 
    \psixLEl = \intfour \xEz\, \intsix \LambdaEz\, 
                  \kersym(\xE-\xEz,\LambdaE\LambdaEz^{-1};\lambda-\lambdaz) 
                  \psixLElz \,.
\end{equation}

Using \eqn{eqn:A5a}, substitute into \eqn{eqn:B1} the Euclidean scalar
kernel (as in \eqn{eqn:A5b}, but with a leading factor of $\mi$) and
the $SO(4)$ kernel (\eqn{eqn:A6}), giving
\begin{multline} \label{eqn:B1a}
    \psixLEl = \sum_{\ell_{A},\ell_{B}}
        \intfour \xEz\, \intsix \LambdaEz\, \\
            \kersym^{(\ell_{A},\ell_{B})}(\xE-\xEz;\lambda-\lambdaz)
            \chi^{(\ell_{A},\ell_{B})}(\LambdaE\LambdaEz^{-1})
            \psixLElz \,,
\end{multline}
where
\begin{equation*}
    \kersym^{(\ell_{A},\ell_{B})}(\xE-\xEz;\lambda-\lambdaz)
        \equiv \mi(2\pi)^{-4}\intfour \pE\, \me^{i\pE\cdot(\xE-\xEz)}
                  \me^{-\mi(\lambda-\lambdaz)
                  (\pE^{2}+m^{2}+\Delta m_{A}^{2}+\Delta m_{B}^{2})}\,.
\end{equation*}
Since the group characters provide a complete set of orthogonal 
functions \cite{weyl50}, the function $\psi(\xEz,\LambdaEz;\lambdaz)$ 
can be expanded as
\begin{equation*}
    \psixLElz = \sum_{\ell_{A},\ell_{B}}
        \chi^{(\ell_{A},\ell_{B})}(\LambdaEz)
        \psi^{(\ell_{A},\ell_{B})}(\xEz;\lambdaz) \,.
\end{equation*}
Substituting this into \eqn{eqn:B1a} and using
\begin{equation*}
     \chi^{(\ell_{A},\ell_{B})}(\LambdaE) = \intsix \LambdaEz\, 
         \chi^{(\ell_{A},\ell_{B})}(\LambdaE\LambdaEz^{-1})
         \chi^{(\ell_{A},\ell_{B})}(\LambdaEz)
\end{equation*}
(see \refcite{weyl50}) gives
\begin{equation*}
    \psixLEl = \sum_{\ell_{A},\ell_{B}}
        \chi^{(\ell_{A},\ell_{B})}(\LambdaE)
        \psi^{(\ell_{A},\ell_{B})}(\xE;\lambda) \,,
\end{equation*}
where
\begin{equation} \label{eqn:B1b}
    \psi^{(\ell_{A},\ell_{B})}(\xE;\lambda)
        = \intfour \xEz\,
          \kersym^{(\ell_{A},\ell_{B})}(\xE-\xEz;\lambda-\lambdaz)
          \psi^{(\ell_{A},\ell_{B})}(\xEz;\lambdaz) \,.
\end{equation}

The general amplitude $\psixLEl$ can thus be expanded into a sum of
terms in the various $SU(2) \times SU(2)$ representations, the
coefficients $\psi^{(\ell_{A},\ell_{B})} (\xE;\lambdaz)$ of which each
evolve separately according to \eqn{eqn:B1b}. As is well known,
reflection symmetry requires that a real particle amplitude must
transform according to a $(\ell,\ell)$ or $(\ell_{A},\ell_{B}) \oplus
(\ell_{B},\ell_{A})$ representation. That is, the amplitude function
$\psixLEl$ must either have the form
\begin{equation*}
    \psixLEl = \chi^{(\ell,\ell)}(\LambdaE)
               \psi^{(\ell,\ell)} (\xE;\lambda)
\end{equation*}
or
\begin{equation*}
    \psixLEl
        = \chi^{(\ell_{A},\ell_{B})}(\LambdaE)
          \psi^{(\ell_{A},\ell_{B})} (\xE;\lambda)
        + \chi^{(\ell_{B},\ell_{A})}(\LambdaE)
          \psi^{(\ell_{B},\ell_{A})} (\xE;\lambda) \,.
\end{equation*}

Assuming one of the above two forms, shift the particle mass to
$m'^{2} = m^{2} + 2\Delta m_{\ell}^{2}$ or $m'^{2} = m^{2} + 2\Delta
m_{\ell_{A}}^{2} + 2\Delta m_{\ell_{B}}^{2}$, so that
\begin{equation*}
    \psixLEl = \intfour \xz, \intsix \Lambdaz\,
                   \chi^{(L)}(\LambdaE\LambdaEz^{-1})
                   \kersym(\xE-\xEz;\lambda-\lambdaz) \psixLElz \,,
\end{equation*}
where $\kersym$ here is (the Euclidean version of) the scalar
propagator of \eqn{eqn:A5b}, but now for the shifted mass $m'$, and
$(L)$ is either $(\ell,\ell)$ or $(\ell_{A},\ell_{B})$. That is, the
full kernel must have the form
\begin{equation} \label{eqn:B1c}
    \kersym^{(L)}(\xE-\xEz,\LambdaE\LambdaEz^{-1};\lambda-\lambdaz)
        = \chi^{(L)}(\LambdaE\LambdaEz^{-1})
          \kersym(\xE-\xEz;\lambda-\lambdaz) \,.
\end{equation}

As is conventional, from now on we will use four-dimensional spinor
indices for the $(1/2,0) \oplus (0,1/2)$ representation and vector
indices (also four dimensional) for the $(1,1)$ representation, rather
than the $SU(2) \times SU(2)$ indices $(\ell_{A},\ell_{B})$ (see, for
example, \refcite{weinberg95}). Let $\DLElpl$ be a matrix
representation of the $SO(4)$ group using such indices. Define
correspondingly indexed amplitude functions by
\begin{equation} \label{eqn:B2} 
    \lpl{\psi}(\xE;\lambda) \equiv \intsix \LambdaE\, \DLElpl \psixLEl
\end{equation}
(note the \emph{double} indexing of $\psi$ here). 

These $\lpl{\psi}$ are the elements of an algebra over the $SO(4)$
group for which, given $\xE$ and $\lambda$, the $\psixLEl$ are the
\emph{components}, ``indexed'' by the group elements $\LambdaE$ (see
Section III.13 of \refcite{weyl50}). The product of two such algebra
elements is (with summation implied over repeated up and down indices)
\begin{equation*}
    \begin{split}
        \hilo{\psi_{1}}{l'}{\lbar}(\xE; \lambda)
        \hilo{\psi_{2}}{\lbar}{l}(\xE; \lambda)
            &= \intsix {\LambdaE}_{1}\, \intsix {\LambdaE}_{2}\,
               \hilo{\Dmat}{l'}{\lbar}({\LambdaE}_{1})
               \hilo{\Dmat}{\lbar}{l}({\LambdaE}_{2})
               \psi_{1}(\xE, {\LambdaE}_{1};\lambda) 
               \psi_{2}(\xE, {\LambdaE}_{2}; \lambda) \\
            &= \intsix \LambdaE\, \DLElpl \intsix {\LambdaE}_{1} \,
               \psi_{1}(\xE, {\LambdaE}_{1}; \lambda) 
               \psi_{2}(\xE, {\LambdaE}_{1}^{-1}\LambdaE; \lambda) \\
            &= \lpl{(\psi_{1}\psi_{2})}(\xE;\lambda) \,,
    \end{split}
\end{equation*}
where the second equality follows after setting ${\LambdaE}_{2} =
{\LambdaE}_{1}^{-1}\LambdaE$ from the invariance of the integration
measure of a Lie group (see, for example, \refcite{warner83}, Section
4.11, and \refcite{weyl50}, Section III.12---this property will be
used regularly in the following), and the product components
$(\psi_{1}\psi_{2})(\xE, \LambdaE; \lambda)$ are defined to be
\begin{equation*}
    (\psi_{1}\psi_{2})(\xE, \LambdaE; \lambda)
        \equiv \intsix \LambdaE'\, \psi_{1}(\xE, \LambdaE'; \lambda)
                   \psi_{2}(\xE, \LambdaE^{\prime -1}\LambdaE; \lambda) \,.
\end{equation*}

Now substitute \eqn{eqn:B1} into \eqn{eqn:B2} to get
\begin{multline*}
    \lpl{\psi}(\xE;\lambda)
        = \intsix \LambdaE\, \intfour \xEz\, \intsix \LambdaEz\, \\
          \DLElpl 
          \kersym(\xE-\xEz,\LambdaE\LambdaEz^{-1};\lambda-\lambdaz)
          \psixLElz \,.
\end{multline*}
Changing variables $\LambdaE \to \LambdaE'\LambdaEz$ then gives
\begin{equation*}
    \begin{split}
        \lpl{\psi}(\xE;\lambda)
            &= \intfour \xz\, 
               \left[ 
                   \intsix \LambdaE'\, 
                   \hilo{\Dmat}{l'}{\lbar}(\LambdaE') 
                   \kersym(\xE-\xz,\LambdaE';\lambda - \lambdaz) 
               \right] \\
            &\qquad\qquad\qquad\qquad\qquad\qquad
               \intsix \LambdaEz\, \hilo{\Dmat}{\lbar}{l}(\LambdaEz)
               \psixLElz \\
            &= \intfour \xz\, 
               \hilo{\kersym}{l'}{\lbar}(\xE-\xz;\lambda-\lambdaz)
               \hilo{\psi}{\lbar}{l}(\xz;\lambdaz) \,,
    \end{split}
\end{equation*}
where the kernel for the algebra elements $\lpl{\psi}(\xE;\lambda)$ is
thus
\begin{equation*}
    \kersymlpl(\xE-\xEz;\lambda-\lambdaz) 
        = \intsix \LambdaE\, \DLElpl
              \kersym(\xE-\xEz,\LambdaE;\lambda-\lambdaz) \,.
\end{equation*}
Substituting \eqn{eqn:B1c} into this, and using the definition of the 
character for a specific representation, $\chi(\LambdaE) \equiv 
\tr(\DLE)$, gives
\begin{equation*}
    \kersymlpl(\xE-\xEz;\lambda-\lambdaz) 
        = \left[
              \intsix \LambdaE\, \DLElpl 
              \hilo{\Dmat}{\lbar}{\lbar}(\LambdaE)
          \right]
          \kersym(\xE-\xEz;\lambda-\lambdaz) \,.
\end{equation*}
Use the orthogonality property
\begin{equation*}
    \intsix \LambdaE\, \DLElpl \lohi{\Dmat}{\lbar'}{\lbar}(\LambdaE)
        = \hilo{\delta}{l'}{\lbar'} \lohi{\delta}{l}{\lbar} \,,
\end{equation*}
where the $SO(4)$ integration measure has been normalized so that $\intsix 
\LambdaE = 1$ (see \refcite{weyl50}, Section 11), to get
\begin{equation} \label{eqn:B3}
    \kersymlpl(\xE-\xEz;\lambda-\lambdaz) 
        = \lpl{\delta}\kersym(\xE-\xEz;\lambda-\lambdaz) \,.
\end{equation}

The $SO(4)$ group propagator is thus simply $\lpl{\delta}$. As
expected, this does not have the same form as would be expected for
the $SO(3,1)$ Lorentz group propagator. However, as argued at the end
of \sect{sect:non-scalar:propagator}, the propagator restricted to the
compact $SO(3)$ subgroup of $SO(3,1)$ \emph{is} expected to have the
same form as for the $SO(3)$ subgroup of $SO(4)$. So we turn now to 
the reduction of $SO(3,1)$ to $SO(3)$.

\section{Spin} \label{sect:non-scalar:spin}

In traditional relativistic quantum mechanics, the Lorentz-group
dependence of non-scalar states is reduced to a rotation representation
that is amenable to interpretation as the intrinsic particle spin.
Since, in the usual approach, physical states are considered to have
on-shell momentum, it is natural to use the 3-momentum as the vector
around which the spin representation is induced, using Wigner's
classic ``little group'' argument \cite{wigner39}.

However, in the spacetime path approach used here, the fundamental
states are not naturally on-shell, rather the on-shell states are
given as the time limits of off-shell states \cite{seidewitz06a}.
Further, there are well-known issues with the localization of on-shell
momentum states \cite{newton49, hegerfeldt74}. Therefore, instead of
assuming on-shell states to start, we will adopt the approach of
\refcite{piron78, horwitz82}, in which the spin representation is
induced about an arbitrary timelike vector. This will allow for a
straightforward generalization of the interpretation obtained in the
spacetime path formalism for the scalar case \cite{seidewitz06a}.

First, define the probability amplitudes $\lpl{\psi}(x;\lambda)$ for a
given Lorentz group representation similarly to the correspondingly
indexed amplitudes for $SO(4)$ representations from
\sect{sect:non-scalar:euclidean}. Corresponding to such amplitudes,
define a \emph{set} of ket vectors $\ketpsi\lol$, with a \emph{single}
Lorentz-group representation index. The $\ketpsi\lol$ define a
\emph{vector bundle} (see, for example, \refcite{frankel97}), of the
same dimension as the Lorentz-group representation, over the
scalar-state Hilbert space.

The basis position states for this vector bundle then have the form
$\ketxll$, such that
\begin{equation*} 
    \lpl{\psi}(x;\lambda) 
        = \gmat^{l'\lbar}\,{}_{\lbar}\innerxlpsi\lol \,,
\end{equation*}
with summation assumed over repeated upper and lower indices and
$\gmat$ being the invariant matrix of a given Lorentz group
representation such that
\begin{equation*} 
    \Dmat\adj \gmat \Dmat = \Dmat \gmat \Dmat\adj = \gmat \,,
\end{equation*}
for any member $\Dmat$ of the representation, where $\Dmat\adj$ is the
Hermitian transpose of the matrix $\Dmat$. For the scalar
representation, $\gmat$ is $1$, for the (Weyl) spinor representation
it is the Dirac matrix $\beta$ and for the vector representation it is
the Minkowski metric $\eta$.

In the following, $\gmat$ will be used (usually implicitly) to
``raise'' and ``lower'' group representation indices. For instance,
\begin{equation*} 
    \lpbraxl \equiv \gmat^{l'l}\,{}\lol\braxl \,,
\end{equation*}
so that
\begin{equation} \label{eqn:C0} 
    \lpl{\psi}(x;\lambda) = \hilp\innerxlpsi\lol \,.
\end{equation}
The states $\ketxll$ are then normalized so that
\begin{equation} \label{eqn:C0d} 
    \hilp\inner{x';\lambda}{x;\lambda}\lol 
        = \lpl{\delta}\, \delta^{4}(x' - x) \,,
\end{equation}
that is, they are orthogonal at equal $\lambda$. 

Consider an arbitrary Lorentz transformation $M$. Since $\psixLl$ is a
scalar, it should transform as $\psipxLl = \psil{M^{-1}x',
M^{-1}\Lambda'}$. In terms of algebra elements,
\begin{equation} \label{eqn:C0e} 
    \begin{split}
        \lpl{\psi'}(x';\lambda) 
            &= \intsix \Lambda'\, \Dmatlpl(\Lambda') 
               \psil{M^{-1}x', M^{-1}\Lambda'} \\
            &= \intsix \Lambda\,  
               \hilo{\Dmat}{l'}{\lbar'}(M) 
               \hilo{\Dmat}{\lbar'}{l}(\Lambda)
               \psil{M^{-1}x', \Lambda} \\
            &= \hilo{\Dmat}{l'}{\lbar'}(M)
               \hilo{\psi}{\lbar'}{l}(M^{-1}x; \lambda) \,.
    \end{split}
\end{equation}

Let $\UL$ denote the unitary operator on Hilbert space corresponding
to the Lorentz transformation $\Lambda$. Then, from \eqn{eqn:C0},
\begin{equation*}
    \lpl{\psi'}(x';\lambda) = \hilp\inner{x';\lambda}{\psi'}\lol \\
                            = \hilp\bral{x'} \UL \ketpsi\lol \,.
\end{equation*}
This and \eqn{eqn:C0e} imply that
\begin{equation*}
    \UL^{-1} \ketll{x'} = \ketllp{\Lambda^{-1}x'}\, 
                                \lpl{[\DL^{-1}]} \,,
\end{equation*}
or
\begin{equation} \label{eqn:C0f} 
    \UL \ketxll = \ketllp{\Lambda x}\, \DLlpl \,.
\end{equation}
Thus, the $\ketxll$ are localized position states that transform
according to a representation of the Lorentz group. 

Now, for any future-pointing, timelike, unit vector $n$ ($n^{2} = -1$
and $n^{0} > 0$) define the standard Lorentz transformation
\begin{equation*}
    L(n) \equiv R(\threen) B(|\threen|) R^{-1}(\threen) \,,
\end{equation*}
where $R(\threevec{n})$ is a rotation that takes the $z$-axis into the
direction of $\threevec{n}$ and $B(|\threevec{n}|)$ is a boost of
velocity $|\threevec{n}|$ in the $z$ direction. Then $n = L(n) e$,
where $e \equiv (1, 0, 0, 0)$.

Define the \emph{Wigner rotation} for $n$ and an arbitrary Lorentz
transformation $\Lambda$ to be
\begin{equation} \label{eqn:C1} 
    \WLn \equiv L(\Lambda n)^{-1} \Lambda L(n) \,,
\end{equation}
such that $\WLn e = e$. That is, $\WLn$ is a member of the
\emph{little group} of transformations that leave $e$ invariant. Since
$e$ is along the time axis, its little group is simply the rotation
group $SO(3)$ of the three space axes.

Substituting the transformation 
\begin{equation*}
    \Lambda = L(\Lambda n) \WLn L(n)^{-1} \,,
\end{equation*}
into \eqn{eqn:C0f} gives
\begin{equation*}
    \UL \ketxll = \ketl{\Lambda x}\lolp\, 
                           \lpl{\left[ 
                               \Dmat \left(
                                   L(\Lambda n) \WLn L(n)^{-1}
                               \right) 
                           \right]} \,.
\end{equation*}
Defining
\begin{equation} \label{eqn:C2} 
    \ketlml{x,n} \equiv \ketxllp \lpl{[\Lmat(n)]} \,,
\end{equation}
where $\Lmat(n) \equiv \Dmat(L(n))$, we see that $\ketlml{x,n}$
transforms under $\UL$ as
\begin{equation} \label{eqn:C3} 
    \UL \ketlml{x,n} 
        = \ketlmlp{\Lambda x, \Lambda n}\, 
          \lpl{\left[
              \Dmat \left( \WLn \right)
          \right]} \,,
\end{equation}
that is, according to the Lorentz representation subgroup given by 
$\Dmat(\WLn)$, which is isomorphic to some representation of the 
rotation group.

The irreducible representations of the rotation group (or, more
exactly, its covering group $SU(2)$) are just the spin
representations, with members given by matrices $\Dsps$, where the
$\sigma$ are spin indices. Let $\ketpsi\los$ be a member of a Hilbert
space vector bundle indexed by spin indices. Then there is a linear,
surjective mapping from $\ketpsi\lol$ to $\ketpsi\los$ given by
\begin{equation*}
    \ketpsi\los = \ketpsi\lol \uls \,,
\end{equation*}
where
\begin{equation} \label{eqn:C4} 
    (\ulspt)^{*} \uls = \hilo{\delta}{\sigma'}{\sigma} \,.
\end{equation}
The isomorphism between the rotation subgroup of the Lorentz
group and the rotation group then implies that, for any
rotation $W$, for all $\ketpsi\lol$,
\begin{equation*}
    \ketpsi\lolp \ulpsp \sps{[D(W)]} = \ketpsi\lolp \lpl{[\Dmat(W)]} \uls
\end{equation*}
(with summation implied over repeated $\sigma$ indices, as well as 
$l$ indices) or
\begin{equation} \label{eqn:C5} 
    \ulpsp \sps{[D(W)]} = \lpl{[\Dmat(W)]} \uls \,,
\end{equation}
where $D(W)$ is the spin representation matrix corresponding to $W$.

Define
\begin{equation} \label{eqn:C6} 
    \ketls{x,n} \equiv \ketlml{x,n}\, \uls \,.
\end{equation}
Substituting from \eqn{eqn:C2} gives
\begin{equation} \label{eqn:C7} 
    \ketls{x,n} = \ketxll\, \uls(n) \,.
\end{equation}
where
\begin{equation} \label{eqn:C8} 
    \uls(n) \equiv \hilo{[\Lmat(n)]}{l}{l'}\,
                                  \ulps \,.
\end{equation}
Then, under a Lorentz transformation $\Lambda$, using \eqns{eqn:C3}
and \eqref{eqn:C5},
\begin{equation*}
    \begin{split}
        \UL \ketls{x,n} 
            &= \ketlmlp{\Lambda x, \Lambda n}\, \lpl{[\Dmat(\WLn)]} \uls \\
            &= \ketlmlp{\Lambda x, \Lambda n}\, \hilo{u}{l'}{\sigma'} 
               \sps{[D(\WLn)]} \\
            &= \ketlsp{\Lambda x, \Lambda n}\, \sps{[D(\WLn)]} \,,
    \end{split}
\end{equation*}
that is, $\ketls{x,n}$ transforms according to the appropriate spin
representation. 

Now consider a past-pointing $n$ ($n^{2} = -1$ and $n^{0} < 0$). In 
this case, $-n$ is future pointing so that $-n = L(-n)e$, or $n = 
L(-n)(-e)$. Taking $L(-n)$ to be the standard Lorentz transformation 
for past-pointing $n$, it is thus possible to construct spin states 
in terms of the future-pointing $-n$. However, since the spacial 
part of $n$ is also reversed in $-n$, it is conventional to consider 
the spin sense reversed, too. Therefore, define
\begin{equation} \label{eqn:C9} 
    \vls(n) \equiv (-1)^{j+\sigma} \hilo{u}{l}{-\sigma}(-n) \,,
\end{equation}
for a spin-$j$ representation, and, for past-pointing $n$, take
\begin{equation*} 
    \ketls{x,n} = \ketxll\, \vls(n) \,.
\end{equation*}
    
The matrices $\uls$ and $\vls$ are the same as the spin coefficient
functions in Weinberg's formalism in the context of traditional field
theory \cite{weinberg64a} (see also Chapter 5 of
\refcite{weinberg95}). Note that, from \eqn{eqn:C5}, using
\eqn{eqn:C1},
\begin{equation*}
    \ulpsp \sps{[D(\WLn)]} 
        = \lpl{[\Dmat(\WLn)]} \uls
        = \lpl{[\Lmat(\Lambda n)^{-1} \DL \Lmat(n)]} \uls \,,
\end{equation*}
so, using \eqn{eqn:C8},
\begin{equation} \label{eqn:C10} 
    \ulsp(\Lambda n) \sps{[D(\WLn)]} = \lpl{[\DL]} \uls(n) \,.
\end{equation}
Using this with \eqn{eqn:C9} gives
\begin{equation*}
    \lpl{[\DL]} \vls(n)
        = (-1)^{\sigma-\sigma'} \vlpsp(\Lambda n)
          \hilo{[D(\WLn)]}{-\sigma'}{-\sigma}\,.
\end{equation*}
Since
\begin{equation*}
    (-1)^{\sigma - \sigma'} \hilo{D(W)}{-\sigma'}{-\sigma}
        = [\sps{D(W)}]^{*}
\end{equation*}
(which can be derived by integrating the infinitesimal case), this
gives,
\begin{equation} \label{eqn:C11} 
    \vlpsp(\Lambda n) [\sps{D(\WLn)}]^{*} 
        = \lpl{[\DL]} \vls(n) \,.
\end{equation}
As shown by Weinberg \cite{weinberg64a, weinberg95}, \eqns{eqn:C10}
and \eqref{eqn:C11} can be used to completely determine the $u$ and
$v$ matrices, along with the usual relationship of the Lorentz group
scalar, spinor and vector representations to the rotation group
spin-0, spin-1/2 and spin-1 representations.

Since, from \eqns{eqn:C4} and \eqref{eqn:C8},
\begin{equation*} 
    \begin{split}
        \ulspt(n)^{*}\uls(n) &= [\lohi{\Lmat(n)}{l}{\lbar'}]^{*}
                                (\lohi{u}{\lbar'}{\sigma'})^{*}
                                \hilo{[\Lmat(n)]}{l}{\lbar}\,
                                \hilo{u}{\lbar}{\sigma} \\
                             &= (\lohi{u}{\lbar'}{\sigma'})^{*}
                                \hilo{[\Lmat(n)^{-1}]}{\lbar'}{l}
                                \hilo{[\Lmat(n)]}{l}{\lbar}\,
                                \hilo{u}{\lbar}{\sigma} \\
                             &= (\lohi{u}{\lbar}{\sigma'})^{*}
                                \hilo{u}{\lbar}{\sigma} \\
                             &= \sps{\delta} \,,
    \end{split}
\end{equation*}
\eqns{eqn:C0d} and \eqref{eqn:C7} give
\begin{equation} \label{eqn:C12a}
    \hisp\inner{x',n;\lambda}{x,n;\lambda}\los
        = \sps{\delta} \delta^{4}(x'-x)
\end{equation}
(and similarly for past-pointing $n$ with $\vls$), so that, for given $n$ 
and $\lambda$, the $\ketls{x,n}$ form an orthogonal basis. However, 
for different $\lambda$, the inner product is
\begin{equation} \label{eqn:C12b}
    \hisp\inner{x,n;\lambda}{\xz,n;\lambdaz}\los
        = \kernelsps \,,
\end{equation}
where $\kernelsps$ is the kernel for the rotation group. As previously
argued, this should have the same form as the Euclidean kernel of
\eqn{eqn:B3}, restricted to the rotation subgroup of $SO(4)$. That is
\begin{equation}
    \kernelsps = \sps{\delta} \kerneld \,.
\end{equation}

As in \eqn{eqn:A1a}, the propagator is given by integrating the 
kernel over $\lambda$:
\begin{equation*}
    \propsps = \sps{\delta} \prop \,,
\end{equation*}
where (using \eqn{eqn:A5b})
\begin{equation*}
    \prop = \int_{\lambdaz}^{\infty} \dl\, \kerneld
          = -\mi(2\pi)^{-4}\intfour p\, 
                \frac{\me^{\mi p\cdot(x-\xz)}}
                     {p^{2}+m^{2}-\mi\epsilon} \,,
\end{equation*}
the usual Feynman propagator \cite{seidewitz06a}. Defining
\begin{equation*}
   \kets{x,n} \equiv \int_{\lambdaz}^{\infty} \dl\, \ketls{x,n}
\end{equation*}
then gives
\begin{equation} \label{eqn:C13}
    \hisp\inner{x,n}{\xz,n;\lambdaz}\los = \propsps \,.
\end{equation}

Finally, we can inject the spin-representation basis states
$\ketls{x,n}$ back into the Lorentz group representation by
\begin{equation*}
    \ketll{x,n} \equiv \ketls{x,n}\ulst(n)^{*} \,,
\end{equation*}
(and similarly for past-pointing $n$ with $\vlst$). Substituting 
\eqn{eqn:C7} into this gives
\begin{equation} \label{eqn:C13a}
    \ketll{x,n} = \ketxllp\Plpl(n) \,,
\end{equation}
where
\begin{equation} \label{eqn:C14}
    \Plpl(n) \equiv \ulps(n)\ulst(n)^{*} = \vls(n)\vlst(n)^{*}
\end{equation}
(the last equality following from \eqn{eqn:C9}). Using 
\eqns{eqn:C12a} and \eqref{eqn:C12b}, the kernel for these states is
\begin{equation*}
    \hilp\inner{x,n;\lambda}{\xz,n;\lambdaz}\lol
        = \Plpl(n) \kerneld \,.
\end{equation*}
However, using \eqns{eqn:C10} and \eqref{eqn:C11}, it can be shown 
that the $\ketll{x,n}$ transform like the $\ketxll$:
\begin{equation*}
    \UL \ketll{x,n} 
        = \ketllp{\Lambda x, \Lambda n}\, \Dmatlpl(\lambda) \,.
\end{equation*}

Taking
\begin{equation*}
    \ket{x,n}\lol \equiv \int_{\lambdaz}^{\infty} \dl\, \ketll{x,n}
\end{equation*}
and using \eqn{eqn:C13} gives the propagator
\begin{equation} \label{eqn:C15}
    \hilp\inner{x,n}{\xz,n;\lambdaz}\lol = \Plpl(n) \prop \,.
\end{equation}
Now, the $\ketll{x,n}$ do not span the full Lorentz group Hilbert
space vector bundle of the $\ketxll$, but they do span the subspace
corresponding to the rotation subgroup. Therefore, using 
\eqn{eqn:C13a} and the idempotency of $\Plpl(n)$ as a projection 
matrix,
\begin{equation} \label{eqn:C16}
    \begin{split}
        \ket{x,n}\lol &= \intfour \xz\, 
                              \hilp\inner{\xz,n;\lambdaz}{x,n}\lol
                              \ketlz{\xz,n}\lolp \\
                      &= \intfour \xz\, \Plpl(n)\prop^{*} 
                              \hilo{P}{\lbar'}{l'}(n) \ketxlz_{\lbar'} \\
                      &= \intfour \xz\, \Plpl(n)\prop^{*} \ketxlzlp \,.
    \end{split}
\end{equation}

\section{Particles and Antiparticles} 
\label{sect:non-scalar:antiparticles}

Because of \eqn{eqn:C13}, the states $\kets{x,n}$ allow for a
straightforward generalization of the treatment of particles and
antiparticles from \refcite{seidewitz06a} to the non-scalar case. As in
that treatment, consider particles to propagate \emph{from} the past
\emph{to} the future while antiparticles propagate from the
\emph{future} into the \emph{past} \cite{stueckelberg41,
stueckelberg42, feynman49}. Therefore, postulate non-scalar particle
states $\ketans{x}$ and antiparticle states $\ketrns{x}$ as follows.

\begin{postulate}
    Normal particle states $\ketans{x}$ are such that
    \begin{equation*}
        \hisp\inner{\adv{x},n}{\xz,n;\lambdaz}\los 
            = \thetaax \propsps
            = \thetaax \propasps \,,
    \end{equation*}
    and antiparticle states $\ketrns{x}$ are such that
    \begin{equation*}
        \hisp\inner{\ret{x},n}{\xz,n;\lambdaz}\los 
            = \thetarx \propsps
            = \thetarx \proprsps \,,
    \end{equation*}
    where $\theta$ is the Heaviside step function, $\theta(x) = 0$, 
    for $x < 0$, and $\theta(x) = 1$, for $x > 0$, and
    \begin{equation*}
        \proparsps = \sps{\delta} (2\pi)^{-3}
                     \intthree p\, (2\Ep)^{-1}
                     \me^{\mi[\mp\Ep(x^{0}-\xz^{0})
                             +\threep\cdot(\threex-\threex_{0})]} \,,
    \end{equation*}
    with $\Ep \equiv \sqrt{\threep^{2} + m^{2}}$. 
\end{postulate}
Note that the vector $n$ used here is timelike but otherwise 
arbitrary, with no commitment that it be, e.g., future-pointing for 
particles and past-pointing for antiparticles.

This division into particle and antiparticle paths depends, of course,
on the choice of a specific coordinate system in which to define the
time coordinate. However, if we take the time limit of the end point
of the path to infinity for particles and negative infinity for
antiparticles, then the particle/antiparticle distinction will be
coordinate system independent.

In taking this time limit, one cannot expect to hold the 3-position of
the path end point constant. However, for a free particle, it is
reasonable to take the particle \emph{3-momentum} as being fixed.
Therefore, consider the state of a particle or antiparticle with a
3-momentum $\threep$ at a certain time $t$.
\begin{postulate}
    The state of a particle ($+$) or antiparticle ($-$) with 
    3-momentum $\threep$ is given by
    \begin{equation*}
        \ketarns{t,\threep}
            \equiv (2\pi)^{-3/2} \intthree x\, 
                    \me^{\mi(\mp\Ep t + \threep\cdot\threex)} 
                    \ketarns{t,\threex} \,.
    \end{equation*}
\end{postulate}

Now, following the derivation in \refcite{seidewitz06a}, but carrying
along the spin indices, gives
\begin{equation} \label{eqn:D1}
    \begin{split}
        \ketans{t,\threep} &=
              (2\Ep)^{-1} \int_{-\infty}^{t} \dt_{0}\,
                     \ketanlzs{t_{0},\threep} \quad \text{and} \\
        \ketrns{t,\threep} &=
              (2\Ep)^{-1} \int_{t}^{+\infty} \dt_{0}\,
                     \ketrnlzs{t_{0},\threep} \,,
    \end{split}
\end{equation}
where
\begin{equation} \label{eqn:D1a}
    \ketarnlzs{t,\threep}
        \equiv (2\pi)^{-3/2} \intthree x\, 
                \me^{\mi(\mp\Ep t + \threep\cdot\threex)} 
                \ketlzs{t,\threex,n} \,.
\end{equation}
Since
\begin{equation*}
    \hisp\inner{\advret{t',\threepp},n; \lambdaz}
          {\advret{t,\threep},n; \lambdaz}\los =
        \sps{\delta} \delta(t'-t) \delta^{3}(\threepp - \threep) \,,
\end{equation*}
we have, from \eqn{eqn:D1},
\begin{equation*}
    \hisp\inner{\advret{t,\threep},n}
               {\advret{t_{0}, \threep_{0}{}},n; \lambdaz}\los =
        (2\Ep)^{-1} \sps{\delta} \theta(\pm(t-t_{0})) 
                    \delta^{3}(\threep - \threep_{0}) \,.
\end{equation*}
Defining the time limit particle and antiparticle states
\begin{equation} \label{eqn:D2}
    \ketarthreepns \equiv \lim_{t \to \pm\infty} 
                                  \ketarns{t,\threep} \,,
\end{equation}
then gives
\begin{equation} \label{eqn:D3}
    \hisp\inner{\advret{\threep},n}
               {\advret{t_{0}, \threep_{0}{},n}; \lambdaz}\los
            = (2\Ep)^{-1} \sps{\delta} 
                    \delta^{3}(\threep - \threep_{0}) \,,
\end{equation}
for \emph{any} value of $t_{0}$. 

Further, writing
\begin{equation*}
    \ketarnlzs{t_{0}, \threep}
        = (2\pi)^{-1/2} \me^{\mp\mi\Ep t_{0}}
                \int \dif p^{0}\, \me^{\mi p^{0}t_{0}} 
                                  \ketlzs{p,n} \,, 
\end{equation*}
where
\begin{equation} \label{eqn:D4}
    \ketlzs{p,n} \equiv (2\pi)^{-2} \intfour x\, \me^{\mi p \cdot x}
                                                 \ketlzs{x,n} 
\end{equation}
is the corresponding 4-momentum state, it is straightforward to see
from \eqn{eqn:D1} that the time limit of \eqn{eqn:D2} is
\begin{equation} \label{eqn:D5}
    \ketarthreepns \equiv \lim_{t \to \pm\infty} \ketarns{t,\threep}
        = (2\pi)^{1/2} (2\Ep)^{-1} \ketarnlzs{\pm\Ep,\threep} \,.
\end{equation}
Thus, a normal particle ($+$) or antiparticle ($-$) that has
3-momentum $\threep$ as $t \to \pm\infty$ is \emph{on-shell}, with
energy $\pm\Ep$. Such on-shell particles are unambiguously normal
particles or antiparticles.

For the on-shell states $\ketarthreepns$, it now becomes reasonable to
introduce the usual convention of taking the on-shell momentum vector
as the spin vector. That is, set $\npar \equiv (\pm\Ep, \threep)/m$
and define 
\begin{equation*}
    \varketarthreep\los \equiv \kets{\advret{\threep},\npar}
\end{equation*}
and
\begin{equation*}
    \varketar{t,\threep}\los \equiv \kets{t,\advret{\threep},\npar} \,,
\end{equation*}
so that
\begin{equation*}
    \varketarthreep\los =
        \lim_{t\to\pm\infty} \varketar{t,\threep}\los \,.
\end{equation*} 
Further, define the position
states
\begin{equation} \label{eqn:D6}
    \begin{split}
        \varketax\lol 
             &\equiv (2\pi)^{-3/2}\intthree p\, 
                 \me^{\mi(\Ep x^{0} - \threep\cdot\threex)}
                 \varketa{x^{0},\threep}\los \ulst(\npa)^{*}
                 \text{ and } \\
         \varketrx\lol 
             &\equiv (2\pi)^{-3/2}\intthree p\,
                 \me^{\mi(-\Ep x^{0} - \threep\cdot\threex)}
                 \varketr{x^{0},\threep}\los \vlst(\npr)^{*} \,.
     \end{split}
\end{equation}
Then, working the previous derivation backwards gives
\begin{equation*}
    \hilp(\advret{x}\ketxlzl = \thetaarx \proparlpl \,,
\end{equation*}
where
\begin{equation*}
    \proparlpl \equiv 
        (2\pi)^{-3} \intthree p\, \Plpl(\npar)
            (2\Ep)^{-1} \me^{\mi[\pm\Ep (x^{0}-\xz^{0}) -
                             \threep\cdot(\threex-\threex_{0})]} \,.
\end{equation*}

Now, it is shown in \refcites{weinberg64a, weinberg95} that the 
covariant non-scalar propagator
\begin{equation*} 
    \proplpl = -\mi(2\pi)^{-4} \intfour p\, \Plpl(p/m)
               \frac{\me^{\mi p\cdot(x-\xz)}}{p^{2}+m^{2}-\mi\varepsilon} \,,
\end{equation*}
in which $\Plpl(p/m)$ has the polynomial form of $\Plpl(n)$, but $p$
is not constrained to be on-shell, can be decomposed into
\begin{equation*}
    \proplpl = \thetaax\propalpl + \thetarx\proprlpl
             + \Qlpl\left(-\mi\pderiv{}{x}\right) 
               \mi\delta^{4}(x-\xz) \,,
\end{equation*}
where the form of $\Qlpl$ depends on any non-linearity of $\Plpl(p/m)$
in $p^{0}$. Then, defining
\begin{equation*}
    \varketx\lol \equiv \intfour \xz\, \proplpl^{*} \ketxlz\lolp \,,
\end{equation*}
$\varketax\lol$ and $\varketrx\lol$ can be considered as a 
particle/antiparticle partitioning of $\varketx\lol$, in a similar 
way as the partitioning of $\ket{x,n}\los$ into $\keta{x,n}\los$ and 
$\ketr{x,n}\los$:
\begin{equation*}
    \begin{split}
        \thetaarx\hilp(x\ketxlzl &= \thetaarx \proplpl \\
                                 &= \thetaarx \proparlpl \\
                                 &= \hilp(\advret{x}\ketxlzl \,.
    \end{split}
\end{equation*}
Because of the delta function, the term in $\Qlpl$ does not 
contribute for $x \neq \xz$.

The states $\ket{x,n}\lol$ and $\varketx\lol$ both transform according
to a representation $\Dmatlpl$ of the Lorentz group, but it is
important to distinguish between them. The $\ket{x,n}\lol$ are
projections back into the Lorentz group of the states $\kets{x,n}$
defined on the rotation subgroup, in which that subgroup is obtained
by uniformly reducing the Lorentz group about the axis given by $n$.
The $\varketx\lol$, on the other hand, are constructed by
inverse-transforming from the momentum states
$\varketar{t,\threep}\los$, with each superposed state defined over a
rotation subgroup reduced along a different on-shell momentum vector.

One can further highlight the relationship of the $\varketx\lol$ to 
the momentum in the position representation by the formal equation
(using \eqn{eqn:C16})
\begin{equation*}
    \varketx\lol = \intfour \xz\, 
                        \Plpl\left( 
                            \mi m^{-1} \pderiv{}{x} 
                        \right)
                        \prop^{*} \ketxlzlp
                 = \ket{x, \mi m^{-1} \partial/\partial x}\lol
                 = \Plpl\left( 
                       \mi m^{-1} \pderiv{}{x} 
                   \right) \ketx\lolp \,.
\end{equation*}
The $\varketx\lol$ correspond to the position states used in
traditional relativistic quantum mechanics, with associated on-shell
momentum states $\varketarthreep$. However, we will see in the next
section that the states $\ket{x,n}\lol$ provide a better basis for
generalizing the scalar probability interpretation discussed in
\refcite{seidewitz06a}.

\section{On-Shell Probability Interpretation} 
\label{sect:non-scalar:probability}

Similarly to the scalar case \cite{seidewitz06a}, let $\HilbH^{(j,n)}$
be the Hilbert space of the $\ketnlzs{x}$ for the spin-$j$
representation of the rotation group and a specific timelike vector
$n$, and let $\HilbH^{(j,n)}_{t}$ be the subspaces spanned by the
$\ketnlzs{t,\threex}$, for each $t$, forming a foliation of
$\HilbH^{(j,n)}$. Now, from \eqn{eqn:D1a}, it is clear that the
particle and antiparticle 3-momentum states $\ketarnlzs{t,\threep}$
also span $\HilbH^{(j,n)}_{t}$. Using these momentum bases, states in
$\HilbH^{(j,n)}_{t}$ have the form
\begin{equation*}
    \ketarnlzs{t, \psi}
        = \intthree p\, \sps{\psi}(\threep) \ketarnlzsp{t, \threep} \,,
\end{equation*}
for matrix functions $\psi$ such that $\tr(\psi\adj\psi)$ is
integrable. Conversely, it follows from \eqn{eqn:D3} that the 
probability amplitude $\sps{\psi}(\threep)$ is given by
\begin{equation} \label{eqn:E0}
    \sps{\psi}(\threep) = (2\Ep)\hisp\inner{\advret{\threep},n}
                                  {\advret{t,\psi},n; \lambdaz}\los \,.
\end{equation}

Let $\HilbH^{\prime (j,n)}_{t}$ be the space of linear functions dual
to $\HilbH^{(j,n)}_{t}$. Via \eqn{eqn:E0}, the bra states
$\his\braathreep$ can be considered as spanning subspaces
$\advret{\HilbH}^{\prime (j,n)}$ of the $\HilbH^{\prime (j,n)}_{t}$,
with states of the form
\begin{equation*}
    \his\bra{\advret{\psi},n}
        = \intthree p\, \lohi{\psi}{\sigma'}{\sigma}(\threep)^{*}\;
                        \hisp\bra{\advret{\threep},n} \,.
\end{equation*}
The inner product
\begin{equation*}
    (\psi_{1},\psi_{2}) 
        \equiv \his\inner{\advret{\psi_{1}{}},n}
               {\advret{t,\psi_{2}{}},n;\lambdaz}\los
        = \int \frac{\dthree p}{2\Ep} 
               \lohi{\psi_{1}}{\sigma'}{\sigma}(\threep)^{*}
               \sps{\psi_{2}}(\threep)
\end{equation*}
gives
\begin{equation*}
    (\psi,\psi)
        = \int \frac{\dthree p}{2\Ep} 
               \sum_{\sigma'\sigma} \sqr{\sps{\psi}(\threep)}
        \geq 0 \,,
\end{equation*}
so that, with this inner product, the $\HilbH^{(j,n)}_{t}$ actually
are Hilbert spaces in their own right.

Further, \eqn{eqn:D3} is a \emph{bi-orthonormality relation} with the
corresponding resolution of the identity (see \refcite{akhiezer81} and
App.\ A.8.1 of \refcite{muynk02})
\begin{equation*} \label{eqn:E1} 
    \intthree p\, 
        (2\Ep) \ketarnlzs{t, \threep}\;\his\bra{\advret{\threep},n} 
            = 1 \,.
\end{equation*}
The operator $(2\Ep) \ketarnlzs{t, \threep}\;\his\braar{\threep,n}$
represents the quantum proposition that an on-shell, non-scalar
particle or antiparticle has 3-momentum $\threep$.

Like the $\lpl{\psi}$ discussed in \sect{sect:non-scalar:euclidean} for
the Lorentz group, the $\sps{\psi}$ form an algebra over the rotation
group with components $\psi(\threep, B)$, where $\Bsps$ is a member
of the appropriate representation of the rotation group, such that
\begin{equation} \label{eqn:E2} 
    \sps{\psi}(\threep)
        = \intthree B\, \Bsps \psi(\threep, B) \,,
\end{equation}
with the integration taken over the 3-dimensional rotation group.
Unlike the Lorentz group, however, components can also be reconstructed
from the $\sps{\psi}(\threep)$ by
\begin{equation} \label{eqn:E3} 
    \psi(\threep, B) = \beta^{-1}\hilo{(B^{-1})}{\sigma}{\sigma'} 
                       \sps{\psi}(\threep) \,
\end{equation}
where
\begin{equation*}
    \beta \equiv \frac{1}{2j+1} \intthree B \,,
\end{equation*}
for a spin-$j$ representation, is finite because the rotation group is
closed. Plugging \eqn{eqn:E3} into the right side of \eqn{eqn:E2}
and evaluating the integral does, indeed, give $\sps{\psi}(\threep)$,
as required, because of the orthogonality property
\begin{equation*} 
    \intthree B\, \Bsps \hilo{(B^{-1})}{\sbar}{\sbar'}
        = \beta \hilo{\delta}{\sigma'}{\sbar'} \lohi{\delta}{\sigma}{\sbar}
\end{equation*}
(see \refcite{weyl50}, Section 11). We can now adjust the group 
volume measure $\dthree B$ so that $\beta = 1$.

The set of all $\psi(\threep, B)$ constructed as in \eqn{eqn:E3} forms
a subalgebra such that each $\psi(\threep, B)$ is uniquely determined
by the corresponding $\sps{\psi}(\threep)$ (see \refcite{weyl50},
pages 167ff). We can then take $\sqr{\psi(\threep, B)} =
\sqr{\hilo{(B^{-1})}{\sigma}{\sigma'}\sps{\psi}(\threep)}$ to be the
probability density for the particle or antiparticle to have
3-momentum $\threep$ and to be rotated as given by $B$ about the axis
given by the spacial part of the unit timelike 4-vector $n$. The
probability density for the particle or antiparticle in 3-momentum
space is
\begin{equation*}
    \intthree B\, \sqr{\psi(\threep, B)}
        = \lohi{\psi}{\sigma'}{\sigma}(\threep)^{*}
          \sps{\psi}(\threep) \,
\end{equation*}
with the normalization
\begin{equation*}
    (\psi,\psi)
        = \int \frac{\dthree p}{2\Ep} 
          \lohi{\psi}{\sigma'}{\sigma}(\threep)^{*}
          \sps{\psi}(\threep) 
        = 1 \,.
\end{equation*}

Next, consider that $\ketnlzs{t,\threex}$ is an eigenstate of the
three-position operator $\op{\threevec{X}}$, representing a particle
localized at the three-position $\threex$ at time $t$. From
\eqn{eqn:E0}, and using the inverse Fourier transform of \eqn{eqn:D4}
with \eqn{eqn:D5}, its three momentum wave function is
\begin{equation} \label{eqn:E4}
    (2\Ep)\, \hisp\inner{\advret{\threep},n}
                     {t,\threex;\lambdaz}\los
        = (2\pi)^{-3/2} \sps{\delta}
          \me^{\mi(\pm\Ep t - \threep\cdot\threex)} \,.
\end{equation}
This is just a plane wave, and it is an eigenfunction of the operator
\begin{equation*}
    \me^{\pm\mi\Ep t} \mi \pderiv{}{\threep} \me^{\mp\mi\Ep t} \,,
\end{equation*}
which acts as the identity on the spin indices and is otherwise the
traditional momentum representation $\mi \partial/\partial\threep$ of
the three-position operator $\op{\threevec{X}}$, translated to time
$t$.

This result exactly parallels that of the scalar case 
\cite{seidewitz06a}. Note that this is only so because of the use of 
the independent vector $n$ for reduction to the rotation group, 
rather than the traditional approach of using the three-momentum 
vector $\threep$. Indeed, it is not even possible to define a 
spin-indexed position eigenstate in the traditional approach, 
because, of course, the momentum is not sharply defined for such a 
state \cite{piron78, horwitz82}.

On the other hand, consider the three-position states $\varketarx\lol$
introduced at the end of \sect{sect:non-scalar:antiparticles}. Even
though these are Lorentz-indexed, they only span the rotation
subgroup. Therefore, we can form their three-momentum wave functions
in the $\his\varbra{\advret{\threep}}$ bases. Using \eqns{eqn:D6} and 
\eqref{eqn:D3},
\begin{equation} \label{eqn:E5}
    (2\Ep)\, \his\varinner{\advret{\threep}}{\advret{x}}\lol 
        = (2\pi)^{-3/2} \ulst(\np)^{*}
          \me^{\mi(\pm\Ep t - \threep\cdot\threex)} \,.
\end{equation}
At $t = 0$, up to normalization factors of powers of $(2\Ep)$, this is
just the Newton-Wigner wave function for a localized particle of
non-zero spin \cite{newton49}. It is an eigenfunction of the position
operator represented as
\begin{equation} \label{eqn:E6}
    \ulpspt(\np)^{*} \me^{\mi\Ep t} \mi \pderiv{}{\threep}
    \me^{-\mi\Ep t} \ulsp(\np)
\end{equation}
for the particle case, with a similar expression using $\vls$ in the 
antiparticle case. Other than the time translation, this is 
essentially the Newton-Wigner position operator for non-zero spin 
\cite{newton49}.

Note that \eqn{eqn:E4} is effectively related to \eqn{eqn:E5} by a 
generalized Foldy-Wouthuysen transformation \cite{foldy50, case54}. 
However, in the present approach it is \eqn{eqn:E4} that is seen to 
be the primary result, with a natural separation of particle and 
antiparticle states and a reasonable non-relativistic limit, just as 
in the scalar case \cite{seidewitz06a}.

\section{Interactions} \label{sect:non-scalar:interactions}

It is now straightforward to extend the formalism to multiparticle
states and introduce interactions, quite analogously to the scalar
case \cite{seidewitz06a}. In order to allow for multiparticle states
with different types of particles, extend the position state of each
individual particle with a \emph{particle type index} $\nbase$, such
that
\begin{equation*}
    \hilp\inner{x',\nbase';\lambda}{x,\nbase;\lambda}\lol
        = \delta^{l'}_{l}\delta^{\nbase'}_{\nbase}\delta^{4}(x'-x) \,.
\end{equation*}
Then, construct a basis for the Fock space of multiparticle states as
sym\-me\-trized/anti\-sym\-me\-trized products of $N$ single particle
states:
\begin{multline*} 
    \ket{\xnliN}\lolN
        \equiv (N!)^{-1/2} \sum_{\text{perms }\Perm} \delta_{\Perm}
        \ket{\xni{\Perm 1};\lambda_{\Perm 1}}\loli{\Perm 1} \cdots \\
        \ket{\xni{\Perm N};\lambda_{\Perm N}}\loli{\Perm N} \,,
\end{multline*}
where the sum is over permutations $\Perm$ of $1,\ldots,N$, and 
$\delta_{\Perm}$ is $+1$ for permutations with an even number of 
interchanges of fermions and $-1$ for an odd number of interchanges.

Define multiparticle states $\ket{\xniN}\lolN$ as similarly
sym\-me\-trized/anti\-sym\-me\-trized products of $\ketx\lol$ states.
Then,
\begin{equation} \label{eqn:F1} 
    \hilpN\inner{\xnpiN}{\seqN{\xnlzi}}\lolN
       = \sum_{\text{perms }\Perm} \delta_{\Perm}
               \prod_{i = 1}^{N}
               \proplplij{\Perm i}{i} \,,
\end{equation}
where each propagator is also implicitly a function of the mass of the
appropriate type of particle. Note that the use of the same parameter
value $\lambdaz$ for the starting point of each particle path is
simply a matter of convenience. The intrinsic length of each particle
path is still integrated over \emph{separately} in $\ket{\xniN}\lolN$,
which is important for obtaining the proper particle propagator
factors in \eqn{eqn:F1}. Nevertheless, by using $\lambdaz$ as a common
starting parameter, we can adopt a similar notation simplification as
in \refcite{seidewitz06a}, defining
\begin{equation*}
    \ket{\xnlziN}\lolN \equiv \ket{\seqN{\xnlzi}}\lolN \,.
\end{equation*}

It is also convenient to introduce the formalism of creation and
annihilation fields for these multiparticle states. Specifically,
define the creation field $\oppsitl(x,\nbase;\lambda)$ by
\begin{equation*}
    \oppsitl(x,\nbase;\lambda)\ket{\xnliN}\lolN 
        = \ket{x,\nbase,\lambda;\xnliN}_{l,\listN{l}} \,,
\end{equation*}
with the corresponding annihilation field $\oppsil(x,\nbase;\lambda)$
having the commutation relation
\begin{equation*}
    [\oppsilp(x',\nbase';\lambda), \oppsitl(x,\nbase;\lambdaz)]_{\mp}
        = \delta^{\nbase'}_{\nbase}\propsymlpl(x'-x;\lambda-\lambdaz) \,,
\end{equation*}
where the upper $-$ is for bosons and the lower $+$ is for fermions.
Further define
\begin{equation*}
    \oppsil(x,\nbase) \equiv 
        \int_{\lambdaz}^{\infty} \dl\, \oppsil(x,\nbase;\lambda) \,,
\end{equation*}
so that
\begin{equation*}
    [\oppsilp(x',\nbase'), \oppsitl(x,\nbase;\lambdaz)]_{\mp}
        = \delta^{\nbase'}_{\nbase}\propsymlpl(x'-x) \,,
\end{equation*}
which is consistent with the multi-particle inner product as given in
\eqn{eqn:F1}. Finally, as in \refcite{seidewitz06a}, define a
\emph{special adjoint} $\oppsi\dadj$ by
\begin{equation} \label{eqn:F2}
    \oppsi\dadj\lol(x,\nbase) = \oppsitl(x,\nbase;\lambdaz) \text{ and }
    \oppsi\dadj\lol(x,\nbase;\lambdaz) = \oppsitl(x,\nbase) \,,
\end{equation}
which allows the commutation relation to be expressed in the more 
symmetric form
\begin{equation*}
    [\oppsilp(x',\nbase'), \oppsi\dadj\lol(x,\nbase)]_{\mp}
        = \delta^{\nbase'}_{\nbase}\propsymlpl(x'-x) \,.
\end{equation*}

We can now readily generalize the postulated interaction vertex
operator of \refcite{seidewitz06a} to the non-scalar case.

\begin{postulate}
    An interaction vertex, possibly occurring at any position in
    spacetime, with some number $a$ of incoming particles and some
    number $b$ of outgoing particles, is represented by the operator
    \begin{equation} \label{eqn:F3} 
        \opV \equiv g\hilpn{a}{}\loln{b} \intfour x\,
                \prod_{i = 1}^{a} \oppsi\dadj_{l'_{i}}(x,\nbase'_{i})
                \prod_{j = 1}^{b} \oppsi^{l_{j}}(x,\nbase_{j}) \,,
    \end{equation}
    where the coefficients $g\hilpn{a}{}\loln{b}$ represent the
    relative probability amplitudes of various combinations of indices
    in the interaction and $\oppsi\dadj$ is the special adjoint
    defined in \eqn{eqn:F2}.
\end{postulate}

Given a vertex operator defined as in \eqn{eqn:F3}, the interacting 
transition amplitude, with any number of intermediate interactions, 
is then
\begin{multline} \label{eqn:F4} 
    G(\xnpiNp | \xniN)\hilpn{N'}{}\lolN \\
        = \hilpn{N'}\bra{\xnpiN} \opG \ket{\xnlziN}\lolN \,,
\end{multline}
where
\begin{equation*}
    \opG \equiv \sum_{m=0}^{\infty} \frac{(-\mi)^{m}}{m!}\opV^{m}
              = \me^{-i\opV} \,.
\end{equation*}
Each term in this sum gives the amplitude for $m$ interactions, 
represented by $m$ applications of $\opV$. The $(m!)^{-1}$ factor 
accounts for all possible permutations of the $m$ identical factors 
of $\opV$.

Clearly, we can also construct on-shell multiparticle states
$\ket{\parnpiN}\lospn{N'}$ and $\ket{\tparnlziN}\losN$ from the
on-shell particle and antiparticle states $\ketarthreep\los$ and
$\ketarlz{t,\threep}\los$. Using these with the operator $\opG$:
\begin{multline} \label{eqn:F5} 
    G(\parnpiN | \parniN)\hispn{N'}{}\losN \\
        \equiv \left[ \prod_{i=1}^{N'} 2\E{\threepp_{i}} \right]
               \hispn{N'}\bra{\parnpiN} \opG \ket{\tparnlziN}\losN \,,
\end{multline}
results in a sum of Feynman diagrams with the given momenta on
external legs. Note that use of the on-shell states requires
specifically identifying external lines as particles and
antiparticles. For each incoming and outgoing particle, $+$ is chosen
if it is a normal particle and $-$ if it is an antiparticle. (Note 
that ``incoming'' and ``outgoing'' here are in terms of the path 
evolution parameter $\lambda$, \emph{not} time.)

The inner products of the on-shell states for individual incoming and
outgoing particles with the off-shell states for interaction vertices
give the proper factors for the external lines of a Feynman diagram.
For example, the on-shell state $\keta{\threepp}\los$ is obtained in
the $+\infty$ time limit and thus represents a \emph{final} (i.e.,
outgoing in \emph{time}) particle. If the external line for this
particle starts at an interaction vertex $x$, then the line
contributes a factor
\begin{equation*}
    (2\Epp) \hisp\inner{\adv{\threepp}}{x;\lambdaz}\lol
        = (2\pi)^{-3/2}
          \me^{\mi(+\Epp x^{0} - \threepp \cdot \threex)} 
          \ulspt(\threepp)^{*} \,.
\end{equation*}
For an incoming particle on an external line ending at an
interaction vertex $x'$, the factor for this line is (assuming 
$x^{\prime 0} > t$)
\begin{equation*}
    (2\Ep) \hilp\inner{x'}{\adv{t,\threep};\lambdaz}\los
        = (2\pi)^{-3/2}
        \me^{\mi(-\Ep x^{\prime 0} + \threep \cdot \threexp)}
          \ulps(\threep) \,.
\end{equation*}
Note that this expression is independent of $t$, so we can take $t \to
-\infty$ and treat the particle as \emph{initial} (i.e., incoming in
time). The factors for antiparticles are similar, but with the time
sense reversed. Thus, the effect is to remove the propagator factors
from external lines, exactly in the sense of the usual LSZ reduction
\cite{lsz55}.

Now, the formulation of \eqn{eqn:F5} is still not that of the usual
scattering matrix, since the incoming state involves initial particles
but final antiparticles, and vice versa for the outgoing state. To
construct the usual scattering matrix, it is necessary to have
multi-particle states that involve either all initial particles and
antiparticles (that is, they are composed of individual asymptotic
particle states that are all consistently for $t \to -\infty$) or all
final particles and antiparticles (with individual asymptotic states
all for $t \to +\infty$). The result is a formulation in terms of the
more familiar scattering operator $\opS$, which can be expanded in a
Dyson series in terms of a time-dependent version $\opV(t)$ of the
interaction operator. The procedure for doing this is exactly
analogous to the scalar case. For details see \refcite{seidewitz06a}.

\section{Conclusion} \label{sect:conclusion}

The extension made here of the scalar spacetime path approach
\cite{seidewitz06a} begins with the argument in \sect{sect:background}
on the form of the path propagator based on Poincar\'e invariance.
This motivates the use of a path integral over the Poincar\'e group,
with both position and Lorentz group variables, for computation of the
non-scalar propagator. Once the difficulty with the non-compactness of
the Lorentz group is overcome, the development for the non-scalar case
is remarkably parallel to the scalar case.

A natural further generalization of the approach, particularly given
its potential application to quantum gravity and cosmology, would be
to consider paths in curved spacetime. Of course, in this case it is
not in general possible to construct a family of parallel paths over
the entire spacetime, as was done in \sect{sect:non-scalar:propagator}. 
Nevertheless, it is still possible to consider infinitesimal
variations along a path corresponding to arbitrary coordinate
transformations. And one can certainly construct a family of
``parallel'' paths at least over any one coordinate patch on the
spacetime manifold. The implications of this for piecing together a
complete path integral will be explored in future work.

Another direction for generalization is to consider massless
particles, leading to a complete spacetime path formulation for
Quantum Electrodynamics. However, as has been shown in previous work
on relativistically parametrized approaches to QED (e.g.,
\refcite{shnerb93}), the resulting gauge symmetries need to be handled
carefully. This will likely be even more so if consideration is
further extended to non-Abelian interactions. Nevertheless, the
spacetime path approach may provide some interesting opportunities for
addressing renormalization issues in these cases \cite{seidewitz06a}.

In any case, the present paper shows that the formalism proposed in
\refcite{seidewitz06a} can naturally include non-scalar particles. This
is, of course, critical if the approach is to be given the
foundational status considered in \refcite{seidewitz06a} and the
cosmological interpretation discussed in \refcite{seidewitz06b}.

    \appendix*

\section{Evaluation of the $SO(4)$ Path Integral}
\label{app:path}

\begin{theorem*}
    Consider the path integral
    \begin{multline*}
        \kersym(\LambdaE\LambdaEz^{-1};\lambda-\lambdaz)
            = \euc{\zeta} \intDsix \ME\,
                \delta^{6}(\ME(\lambda)\LambdaE^{-1}-I)
                \delta^{6}(\ME(\lambdaz)\LambdaEz^{-1}-I) \\
                \exp \left[
                    \mi\int^{\lambda}_{\lambdaz}\dl'\,
                        \frac{1}{2} \tr(\OmegaE(\lambda')
                                        \OmegaE(\lambda')\T)
                \right]
    \end{multline*}
    over the six dimensional group $SO(4) \sim SU(2) \times SU(2)$,
    where $\OmegaE(\lambda')$ is the element of the Lie algebra $so(4)$
    tangent to the path $\ME(\lambda)$ at $\lambda'$. This path integral
    may be evaluated to get
    \begin{multline} \label{eqn:A1A}
        \kersym(\LambdaE\LambdaEz^{-1};\lambda-\lambdaz) \\
            = \sum_{\ell_{A},\ell_{B}}
                \exp^{-\mi( \Delta m_{\ell_{A}}^{2} 
                          + \Delta m_{\ell_{B}}^{2})
                          (\lambda - \lambdaz)}
                (2\ell_{A}+1)(2\ell_{B}+1)
                \chi^{(\ell_{A}\ell_{B})}(\LambdaE\LambdaEz^{-1}) \,,
    \end{multline}
    where the summation over $\ell_{A}$ and $\ell_{B}$ is from $0$ to
    $\infty$ in steps of $1/2$, $\Delta m_{\ell}^{2} = \ell(\ell+1)$
    and $\chi^{(\ell_{A},\ell_{B})}$ is the group character for the
    $(\ell_{A},\ell_{B})$ $SU(2) \times SU(2)$ group representation.
\end{theorem*}
\begin{proof}
    Parametrize a group element $\ME$ by a six-vector $\theta$ such
    that
    \begin{equation*}
        \ME = \exp(\sum_{i=1}^{6}\theta_{i}J_{i}) \,,
    \end{equation*}
    where the $J_{i}$ are $so(4)$ generators for $SO(4)$. Then
    $\tr(\OmegaE\OmegaE\T) = \dot{\theta}^{2}$, where the dot denotes
    differentiation with respect to $\lambda$. Dividing the six
    generators $J_{i}$ into two sets of three $SU(2)$ generators, the
    six-vector $\theta$ may be divided into two three-vectors
    $\theta_{A}$ and $\theta_{B}$, parametrizing the two $SU(2)$
    subgroups. The path integral then factors into two path integrals
    over $SU(2)$:
    \begin{multline*}
        \kersym(\LambdaE\LambdaEz^{-1};\lambda-\lambdaz) \\
            = \euc{\zeta}^{1/2} \intDthree W_{A}\,
                \delta^{3}(W_{A}(\lambda)B_{A}^{-1}-I)
                \delta^{6}(W_{A}(\lambdaz)B_{A0}^{-1}-I)
                \exp \left[
                    \mi\int^{\lambda}_{\lambdaz}\dl'\,
                        \frac{1}{2} \dot{\theta_{A}}^{2})
                \right] \\
            \times \euc{\zeta}^{1/2} \intDthree W_{B}\,
                \delta^{3}(W_{B}(\lambda)B_{B}^{-1}-I)
                \delta^{6}(W_{B}(\lambdaz)B_{B0}^{-1}-I)
                \exp \left[
                    \mi\int^{\lambda}_{\lambdaz}\dl'\,
                        \frac{1}{2} \dot{\theta_{B}}^{2})
                \right] \,,
    \end{multline*}
    where $\LambdaE = B_{A} \otimes B_{B}$ and $\LambdaEz = B_{A0} 
    \otimes B_{B0}$.
    
    The $SU(2)$ path integrals may be computed by expanding the 
    exponential in group characters \cite{kleinert06,bohm87}. The 
    result is
    \begin{multline} \label{eqn:A2A}
        \euc{\zeta}^{1/2} \intDthree W\,
                    \delta^{3}(W(\lambda)B^{-1}-I)
                    \delta^{6}(W(\lambdaz)B_{0}^{-1}-I)
                    \exp \left[
                        \mi\int^{\lambda}_{\lambdaz}\dl'\,
                            \frac{1}{2} \dot{\theta}^{2})
                    \right] \\
            = \sum_{\ell}\exp^{-\mi\Delta m_{\ell}^{2}
                                  (\lambda - \lambdaz)}
                            (2\ell+1)
                            \chi^{(\ell)}(B B_{0}^{-1}) \,,
    \end{multline}
    where $\chi^{(\ell)}$ is the character for the spin-$\ell$
    representation of $SU(2)$ and the result includes the correction
    for integration ``on'' the group space, as given by Kleinert
    \cite{kleinert06}. The full $SO(4)$ path integral is then given by
    the product of the two factors of the form \eqn{eqn:A2A}, which is
    just \eqn{eqn:A1A}, since \cite{weyl50}
    \begin{equation*}
        \chi^{(\ell_{A},\ell_{B})}(\LambdaE\LambdaEz^{-1}) =
            \chi^{(\ell_{A})}(B_{A}B_{A0}^{-1})
            \chi^{(\ell_{B})}(B_{B}B_{B0}^{-1}) \,.
    \end{equation*}
\end{proof}

    \bibliography{../../RQMbib}
    
\end{document}